\PassOptionsToPackage{lowtilde}{url} 
\documentclass[twocolumn]{aastex631}

\usepackage{graphicx}
\usepackage{amssymb}
\usepackage[fleqn]{amsmath}
\usepackage{times}
\usepackage{subfigure}
\usepackage[T1]{fontenc}
\usepackage{ae,aecompl}
\usepackage{textcomp}

\newcommand{\msun}{$M_{\odot}$}
\newcommand{\lcdm}{$\rm{\Lambda}$-CDM}
\newcommand{\densu}{$\rm{M_{\odot}/kpc^3}$}

\begin{document}

\title{Cosmological predictions for minor axis stellar density profiles in the inner regions of Milky Way-mass galaxies }

\author[0000-0001-7297-8508]{Madeline Lucey}
\affiliation{
Department of Physics \& Astronomy, University of Pennsylvania, 209 S 33rd St., Philadelphia, PA 19104, USA }

\author[0000-0003-3939-3297]{Robyn E. Sanderson}
\affiliation{
Department of Physics \& Astronomy, University of Pennsylvania, 209 S 33rd St., Philadelphia, PA 19104, USA }

\author[0000-0003-1856-2151]{Danny Horta}
\affiliation{Center for Computational Astrophysics, Flatiron Institute, 162 5th Avenue, New York, NY 10010, USA}

\author[0000-0001-8746-4753]{Aritra Kundu}
\affiliation{
Department of Physics \& Astronomy, University of Pennsylvania, 209 S 33rd St., Philadelphia, PA 19104, USA }

\author[0000-0003-3729-1684]{Philip F. Hopkins}
\affiliation{
TAPIR, Mailcode 350-17, California Institute of Technology, Pasadena, CA 91125, USA}

\author[0000-0002-8354-7356]{Arpit Arora}
\affiliation{
Department of Physics \& Astronomy, University of Pennsylvania, 209 S 33rd St., Philadelphia, PA 19104, USA }

\author[0009-0008-8901-2206]{Jasjeev Singh}
\affiliation{
Department of Physics \& Astronomy, University of Pennsylvania, 209 S 33rd St., Philadelphia, PA 19104, USA }

\author[0000-0001-5214-8822]{Nondh Panithanpaisal}
\affiliation{Observatories of the Carnegie Institution for Science, 813 Santa Barbara St, Pasadena, CA 91101, USA}
\affiliation{
TAPIR, Mailcode 350-17, California Institute of Technology, Pasadena, CA 91125, USA}



\begin{abstract}
\lcdm\ cosmology predicts the hierarchical formation of galaxies which build up mass by merger events and accreting smaller systems. The stellar halo of the Milky Way has proven to be a useful tool for tracing this accretion history. However, most of this work has focused on the outer halo where dynamical times are large and the dynamical properties of accreted systems are preserved. In this work, we investigate the inner galaxy regime, where dynamical times are relatively small and systems are generally completely phase-mixed. Using the FIRE-2 and Auriga cosmological zoom-in simulation suites of Milky Way-mass galaxies, we find the stellar density profiles along the minor axis (perpendicular to the galactic disk) within the NFW scale radii (R$\approx$15 kpc) are best described as an exponential disk with scale height <0.3 kpc and a power law component with slope $\alpha\approx$-4. The stellar density amplitude and slope for the power law component is not significantly correlated with metrics of the galaxy's accretion history. Instead, we find the stellar profiles strongly correlate with the dark matter profile. Across simulation suites, the galaxies studied in this work have a stellar to dark matter mass ratio that decreases as $1/r^2$ along the minor axis.
\end{abstract}



\section{Introduction} \label{sec:intro}

Understanding the formation and evolution of galaxies is one of the main goals of astrophysics today. In the inside-out theory of galaxy formation, the innermost regions of galaxies form first and are therefore especially informative for studying the earliest epochs of galaxy formation \citep{Peebles1969,Larson1976,Fall1980,Mo1998,Somerville2008,Dutton2011}. Furthermore, a substantial fraction of a galaxy's stellar mass is within the inner region, making it information-rich. The Milky Way (MW) presents a unique opportunity to study the inner region of a galaxy in exquisite detail. 

Along with more than a third of massive ($\rm{M_*>10^{10}~M_{\odot}}$) disk galaxies in the local Universe \citep{Sellwood1993,Masters2011,Gavazzi2015}, the MW hosts a galactic bar in its center \citep{Blitz1991,Weiland1994,Peters1975,Binney1991}. The bulk of the stellar mass in the inner Galaxy participates in the bar structure \citep{Howard2009, Shen2010, Ness2013b, Debattista2017}. The MW also has an X-shaped structure in its center \citep{Nataf2010,McWilliam2010,Ness2012,Wegg2013,Ness2016b}, which is characteristic of a boxy/peanut-shaped (B/P) bulge and consistent with simulations and observations of barred galaxies \citep{Combes1990,Athanassoula2005,Martinez-Valpuesta2006,Bureau2006,Laurikainen2014,Debattista2019}. 

It is currently debated whether the MW also hosts a less-massive metal-poor classical bulge component \citep{Babusiaux2010,Hill2011,Zoccali2014}. The major evidence for a metal-poor classical bulge is based on the stellar kinematics as a function of metallicity. Specifically, metal-poor stars in the inner Galaxy rotate slower and have a higher velocity dispersion than the metal-rich stars \citep{Ness2013a,Kunder2016,Arentsen2020}. However, \citet{Debattista2017} demonstrated that
these observations may be the result of the overlapping Galactic halo whose density would peak in the center of the Galaxy. In fact, 25\% of the RR Lyrae stars in the inner Galaxy were found on orbits with apocenters $>$3.5 kpc from the Galactic center \citep{Kunder2020}. Similarly, \citet{Lucey2021} found that about 50\% of metal-poor giants in the inner Galaxy are interlopers with apocenters $>$3.5 kpc and that the fraction of interlopers increases with decreasing metallicity. After removing these stars from the sample, \citet{Lucey2021} found that the velocity dispersion decreased and there was no longer evidence for a classical bulge component in the kinematics. These results are supported by the larger Pristine Inner Galaxy Survey \citep{Ardern-Arentsen2024}.

Whether they comprise a classical bulge or the innermost part of the Galactic halo, the stars which are not part of the bar or B/P bulge in the inner Galaxy give us a unique clue to the MW's formation history and galaxy evolution as a whole \citep{Rix2022,Horta2021}. Classical bulges and halos are both thought to be made through mergers and galaxy accretion events \citep{Kauffmann1993,Kobayashi2011a,Guedes2013,Freeman2002,Belokurov2013,Bland-Hawthorn2016,Johnston1995,Helmi2020}. Therefore, it is thought that the total stellar mass and radial profile of the halo/bulge components may reflect the galaxy's cumulative accretion history \citep{Helmi2020,Han2022,Deason2013}.

However, there are also suggestions that the inner halo (Galactocentric radius <10 kpc) may be a distinct Galactic component with in-situ origins. One theory for creating an in-situ inner halo is ELS-like contraction \citep*{Eggen1962}, with early star formation occurring on halo-like random orbits \citep{Carollo2007,Carollo2010,Belokurov2022,El-Badry2018b,Yu2023}. Another proposed theory for the creation of an in-situ inner halo includes disruption of an old thick disk during a major merger \citep{Belokurov2020}. Additionally, it is thought that the inner halo may be composed of disrupted globular clusters \citep{Schiavon2017,Horta2021, Belokurov2024}. By comparing MW observations to cosmological zoom-in simulations, we can better understand the history of these stars, and the MW's evolution.

The radial density profile of the MW's halo is typically quantified with a double power law, although single and triple power laws have also been suggested. Studies use a variety of stellar distance tracers, and radial ranges resulting in large spread in the MW estimates. Generally, the breaking radius estimates range from $\approx$10-30 kpc \citep{ Sesar2013,Han2022} with inner slopes ranging from -1 to -3 \citep{Sesar2010, Sesar2013, Faccioli2014,Han2022,Lane2023, Horta2024} and outer slopes from -3 to -6 \citep{Sesar2010,Sesar2013,Deason2011}. Most of these works are based on stars with Galactocentric radii $>$ 5 kpc. The extension of the stellar density profile within 5 kpc, where Galactic extinction and crowding make observations difficult, is less certain. However, the available estimates are generally consistent with the results at larger radii \citep{Pietrukowicz2015,PerezVillegas2017,Yang2022}. The MW's halo density profile is generally consistent with nearby MW-mass disk galaxies which have power law slopes of -3 to -5 \citep{Harmsen2017}. Furthermore, our nearest disk galaxy neighbor, M31, has a power law slope of -3.7 \citep{Ibata2014}.

Cosmological zoom-in simulations provide a crucial tool for interpreting the observational properties of the MW and other galaxies. Using the IllustrisTNG suite of simulations, \citet{Pillepich2018} found that less massive galaxies generally have steeper halo density slopes, and that the profiles flatten towards the galactic center. They also found MW-mass galaxies to have a stellar density power law slope of $\approx$-4.3 on average for stars within the half-light radius. Using the Auriga simulations of MW-mass galaxies, \citet{Monachesi2019} found that galaxies with fewer progenitors have more massive halos. Furthermore, they found that galaxies with fewer progenitors have steeper halo stellar density profile slopes, but the correlation is quite weak \citep[see Figure 11 in][]{Monachesi2019}. It has also been suggested that the halo stellar density slope is related to the accretion time of the dominant progenitor \citep{D'Souza2018}. In general, results between simulation suites differ, and a definitive metric that predicts the stellar halo density slope is yet to be found.

Similar to MW estimates, results from \lcdm\ simulations of Milky Way-mass galaxies show a large range of radial density profile power law slopes. Whether the works use single or broken power laws, the slopes range from $\approx$ -2 to $\approx$ -6.5 \citep{Monachesi2019,Deason2013,Font2020,Cooper2010,Amorisco2017}, which is consistent with the range observed in nearby disk galaxies, including M31 and the MW \citep{Harmsen2017}. However, studies of the simulated stellar halos use either a kinematic or spatial selection to define the halo which can impact the results \citep{Monachesi2019,Font2020,Pillepich2018,Cooper2010}. For comparison with observations, a spatial selection is generally preferred since kinematic information is not always available. However, in order to avoid contamination by disk stars many of these works only study the halo profile beyond $\sim$ 10 kpc from the galactic center \citep{Monachesi2019,Pillepich2018,Cooper2010}. As we are interested in the inner most region in this work, we instead study the minor axis profile and simultaneously fit a  power-law halo with an exponential disk.

In this work, we use the FIRE-2 and Auriga MW-mass galaxy simulation suites to study inner minor axis stellar density profiles and explore their relation to the galaxy's mass assembly history. In Section \ref{sec:sims} and \ref{sec:asims}, we provide a brief description of the FIRE-2 and Auriga simulation suites, respectively. Our method for parametrizing the density profiles of the simulated galaxies is described in Section \ref{sec:prof} and Appendix \ref{app:a} in more detail. We compare the stellar density profile parameters with metrics of the galaxy's merger and accretion history in Section \ref{sec:not_accr} and with the dark matter profile parameters in Section \ref{sec:dm}. Last in Section \ref{sec:conclu}, we discuss the results and summarize the conclusions. 

\section{Cosmological Zoom-in Simulations}\label{sec:sims}

This work is primarily based on the FIRE-2 cosmological zoom-in simulations of Milky Way-mass galaxies \citep{Wetzel2023}. However, we also make use of the publicly available Auriga simulation suite as comparison \citep{Grand2024}. As these two suites are numerically relatively similar with a few key differences in their physical prescriptions, they provide a unique opportunity to learn about galaxy formation physics by studying how the galaxies differ, or stay the same, between suites.

\subsection{FIRE-2 Milky Way-mass Simulations}

\begin{figure*}
    \centering
    \includegraphics[width=\linewidth]{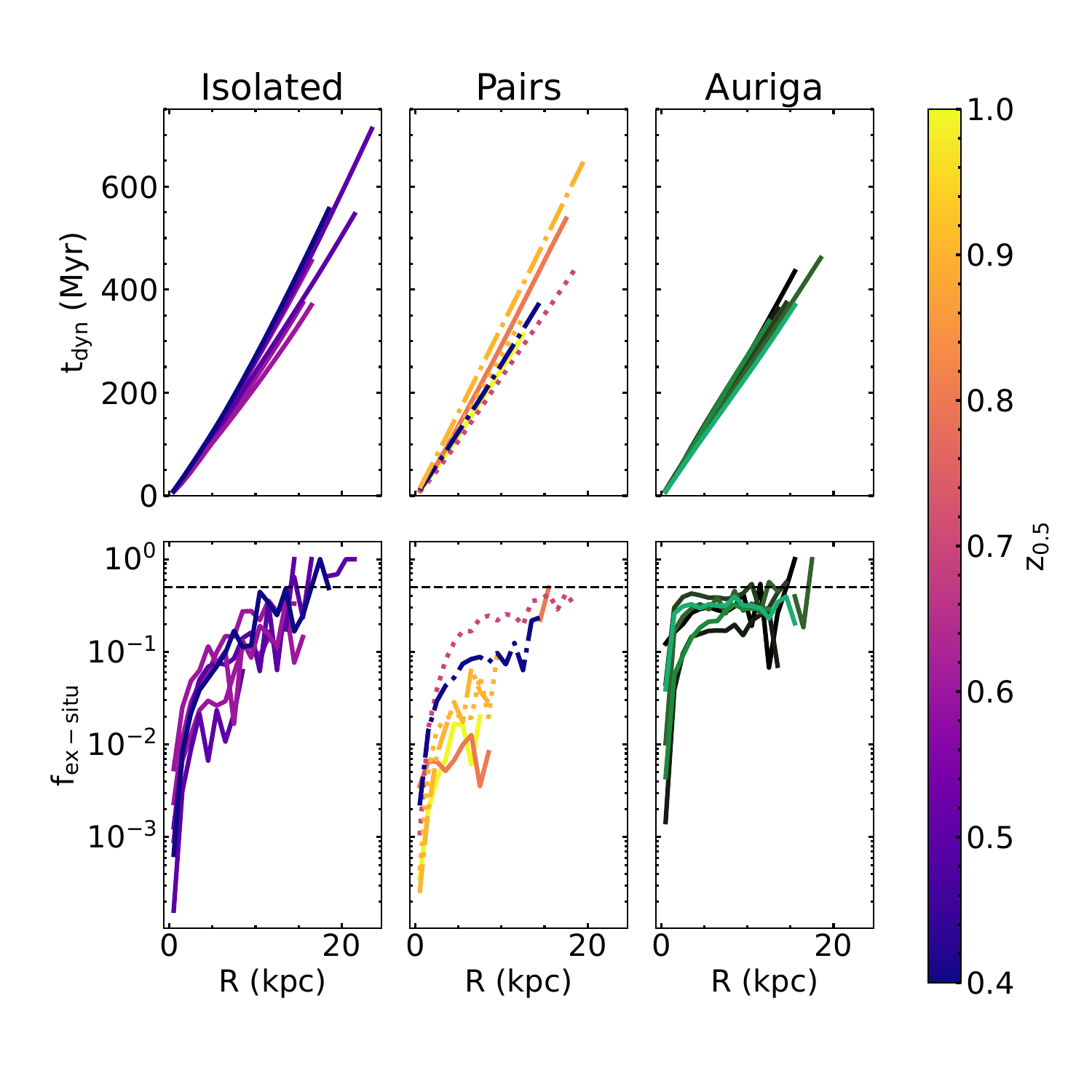}
    \caption{In the top panels we show the dynamical time as a function of galactocentric radius ($R$) from $R$=0 kpc to their respecitve NFW scale radii. The dynamical time is calculated from the mean density of dark matter, stars and gas inside radius $R$. In the lower panels, we show the fraction of accreted stellar mass profile along the minor axis for the same radial range. The left panel shows results for the isolated FIRE-2 simulations while the middle panel shows results for the pairs. Results for these galaxies are colored based on their formation redshift ($\rm{z_{0.5}}$). The right panel shows results for the Auriga galaxies in random shades of green.}
    \label{fig:tdyn}
\end{figure*}

From FIRE-2, we use the \textit{Latte} suite of seven isolated Milky Way-mass galaxies \citep{Wetzel2016} as well as the \textit{ELVIS} suite of three Local Group-like pairs of Milky Way-mass galaxies \citep{Garrison-Kimmel2019}.  All of these simulations are run with the FIRE-2 physics model \citep{Hopkins2018b} using the GIZMO\footnote{\url{http://www.tapir.caltech.edu/~phopkins/Site/GIZMO.html}} gravity plus hydrodynamics code in meshless finite-mass (MFM) mode  \citep{Hopkins2015}. For a complete and detailed description of the simulation implementations, we refer the reader to the above papers. Below, we summarize a few key properties.

Each simulation assumes flat \lcdm\ cosmology with parameters consistent with \citet{Planck2014}. Specifically, the Latte suite (excluding m12w) uses $\Omega_m$ = 0.272, $\Omega_b$ = 0.0455, $\sigma_8$ = 0.807, $n_s$ = 0.961, $h$ = 0.702. Of the Elvis suite of galaxy pairs, Thelma \& Louise and Romulus \& Remus both use the same cosmology as in the original ELVIS dark-matter-only (DMO) suite: $\Omega_m$ = 0.266, $\Omega_b$ = 0.0449, $\sigma_8$ = 0.801, $n_s$ = 0.963, $h$ = 0.71. The third ELVIS galaxy pair, Romeo \& Juliet along with one Latte galaxy, m12w, both use the updated parameters from \citet{Planck2020}, $\Omega_m$ = 0.31, $\Omega_b$ = 0.048, $\sigma_8$ = 0.82, $n_s$ = 0.97, $h$ = 0.68.

The physical prescription for feedback mechanisms are known to significantly impact the distribution of mass in simulated galaxies \citep{Pontzen2014,Lazar2020}. FIRE-2 simulations include implementations of stellar feedback from stellar winds, radiation pressure from young stars, Type II and Type Ia supernovae, photoelectric heating, and photo-heating from ionizing radiation, which regulates star formation. The gas density threshold for star formation is $n_{SF}>1000~\rm{cm^{-3}}$. Feedback event rates, luminosities, energies, mass-loss rates, and other quantities are tabulated directly from stellar evolution models \citep[STARBURST99; ][]{Leitherer1999}. All 13 simulated galaxies have dark matter halo masses at $z$=0 of $\rm{M_{200}= 1-2.1 \times 10^{12}~M_{\odot} }$ \citep{Sanderson2018}. The \textit{Latte} suite galaxies have initial stellar particle masses of 7070 \msun, while the \textit{ELVIS} suite galaxies have higher resolution at initial stellar particle masses of 3500 \msun. Star particle softening lengths are $\approx$ 4 pc and dark matter force softening is $\approx$ 40 pc. 

These simulated galaxies show agreement with the observed stellar mass-dark matter halo mass relation across cosmic time \citep{Hopkins2018b}. They are also consistent with a number of key observed properties of the Milky Way, including the stellar halo mass fraction \citep{Sanderson2018}, the existence of a metal-rich in-situ stellar halo component \citep{Bonaca2017}, and the radial and vertical structure of the stellar disk \citep{Ma2017,Sanderson2020,Bellardini2021,McCluskey2024}. The simulated satellite populations are also consistent with observations of these populations around the Milky Way and M31 \citep{Wetzel2016,Samuel2020,Garrison-Kimmel2019b,Panithanpaisal2021,Cunningham2021}. However, there are also a number of works which have found inconsistencies between the FIRE-2 simulations and Milky Way data. For example, \citet{Shipp2023} finds that while the stellar masses and number of observed streams are consistent with Milky Way observations, the orbital properties of the streams are not. Furthermore, the disks of FIRE-2 galaxies from the Latte suite emerge later and are therefore younger than estimates of the age of the Milky Way disk \citep{McCluskey2024}.

To understand the role of hierarchical formation in building the minor axis stellar profile, we identify and track star particles that formed outside of the main progenitor galaxy across the simulation as a function of time. These star particles are formed in subhalos before they interact with the main branch progenitor. The redshift when the main progenitor reaches a mass that is 3 times larger than the next most massive luminous halo, $\rm{z_{MR_{3:1}}}$, is when the main progenitor emerges as the dominant host galaxy \citep{Santistevan2020,Horta2024}. Before this redshift, systems that merge with the main progenitor are labeled as building blocks \citep{Horta2024}. After this redshift, the luminous halos that are within the virial radius of the central host galaxy at $z=0$ are labeled as accreted. These systems are tracked with the help of the ROCKSTAR halo catalogs and the halo merger trees \citep{Behroozi2013, Behroozi_etal_2013b}; along with every star particle that make-up a present-day substructure (satellites, streams, or phase-mixed) within this virial radius. These accreted substructures are classified as phase-mixed, if they satisfy the following criteria at the present-day \citep{Panithanpaisal2021}: (1) the total stellar mass is greater than $\rm{10^{4.5}}$ $\rm{M_\odot}$, (2) the maximum separation between any two star particles is greater than 120 kpc, and (3) the median of the local velocity dispersion of the star particles is greater than a stellar-mass dependent threshold value (see Equation 2 of \citet{Panithanpaisal2021}).

\subsection{Auriga Simulations} \label{sec:asims}

The Auriga cosmological zoom-in simulations provide a good comparison to the FIRE-2 simulations because they both have similar resolution and broadly MW-like  $z=0$ disk galaxies, but different physical prescriptions.  In this work, we specifically use the 6 Milky Way-mass galaxies simulated at ``level 3" resolution, which is most similar to the resolution of FIRE-2. For a detailed description of these simulations we refer the reader to \citet{Grand2017} and \citet{Grand2024}. Below, we provide a brief overview of a few key details.

The Auriga simulations assume flat \lcdm\ cosmology with parameters taken from \citet{Planck2014}. They are run using the gravo-magnetohydrodynamics moving-mesh code AREPO \citep{Springel2010,Pakmor2016}. The halos are selected from the dark-matter-only EAGLE simulations \citep{Schaye2015} as Milky Way-mass ($1<M_{200}/[10^{12}~M_{\odot}]<2$) halos at $z=0$.  The dark matter particle mass for these simulations is 50,000 \msun\ while the stellar particles have mass 6,000 \msun. The softening lengths for the gas, stars, and dark matter are $\approx$ 188 pc.

The Auriga simulations implement stellar feedback from Type II supernovae. However, unlike the FIRE-2 supernova rates that evolve in time with the star particles, the Auriga simulations have instantaneous feedback at the time of star formation. Furthermore, the Auriga simulations are missing the early stellar feedback prescriptions included in FIRE-2 which are thought to regulate star formation. The gas density threshold for star formation is $n_{SF}>0.13~\rm{cm^{-3}}$, which is significantly smaller than the FIRE-2 threshold of $n_{SF}>1000~\rm{cm^{-3}}$. On the other hand, Auriga includes AGN feedback, while FIRE-2 does not.

For the Auriga galaxies, stellar particles are labeled as accreted if they form in a subhalo and are within the $R_{200}$ of the main host galaxy as $z=0$. Similar to FIRE-2, the Auriga simulations have been shown to produce galaxies that are disk-dominated with Milky Way-mass stellar masses, sizes, rotation curves, and metallicities \citep{Grand2017}. 

\subsection{Inner Galaxy Dynamical Time and Accreted Fraction Profile along the Minor Axis}

To confirm that the inner regions of the simulated galaxies are similarly well phase-mixed, we investigate the dynamical time as a function of galactocentric radius. In the top panels of Figure \ref{fig:tdyn}, we show the dynamical time as a function of galactocentric radius out to the NFW scale radius of the main dark matter halo for each simulated galaxy used in this work. Specifically, we show results for the isolated FIRE-2 galaxies in the left panel, the FIRE-2 pair galaxies in the middle panel, and the Auriga galaxies in the right panel. The FIRE-2 galaxies are colored by the redshift where they have reached 50\% of their $z=0$ mass ($z_{0.5}$). Since the $z_{0.5}$ metric is not available to us for the Auriga galaxies, they are shown in random shades of green. The dynamical times are calculated using the following formula:

\begin{equation}
t_{dyn}(R)= \sqrt{\frac{3\pi}{G\overline{\rho(<R)}}}
\end{equation}
where $G$ is the gravitational constant and $\overline{\rho(<R)}$ is the average density inside radius $R$. As the average density of a galaxy decreases as a function of radius, the dynamical time increases. At the NFW scale radii of these simulated galaxies the dynamical times is on average $\approx$ 500 Myr with a range of $\approx$250-750 Myr. As the phase-mixing timescale is a few times the dynamical time, we expect mergers with infall times $\gtrapprox$ 2 Gyr ago to be well phase-mixed inside this radius.

In the bottom panels of Figure \ref{fig:tdyn}, we explore the birth origin of the stars that comprise the minor axis profile. As the halos/bulges of galaxies are thought to be built from accreted systems, we probe the distance along the minor axis  from the galactic center  where accreted stars begin to outnumber stars that formed in-situ. Specifically, we show the fraction of accreted stars within a cylinder along the minor axis which is defined to be perpendicular to the galaxy's disk plane. The cylinder's radius is $1/\sqrt{\pi}$ so that each cylindrical bin with height of 1 kpc has the volume of 1 $\rm{kpc^3}$. Similar to the top panels, we only calculate the accreted fraction out to the NFW scale radius of each galaxy. The black dashed line indicates the accreted fraction value of 50\%. 

The FIRE-2 galaxies generally increase their accreted fraction with increasing distance from the Galactic center. The isolated galaxies (left panel) typically reach an accreted fraction of $\sim$50\% just inside of the NFW scale radius, while the accreted fraction is generally <10\% within R< 5 kpc.  The paired galaxies (middle panel) show larger spread in their accreted-fraction profiles, with the earlier forming galaxies (larger $z_{0.5}$) having lower accreted fractions closer to the galactic center. The Auriga galaxies show significantly different behavior than the FIRE-2 galaxies in that their accreted fraction quickly jumps to $>$10-40\% at $\sim$2 kpc above the galactic center, rather than the gradual increase seen in FIRE-2. In the Auriga galaxies, the accreted fraction only increases slightly with increasing distance from the galactic center so that just within the NFW scale radii (R$\sim$15 kpc) the galaxies generally have an accreted fraction within 10-50\%. It is possible this may be due to differences in method of defining accreted stars. Both simulations use halo tracking methods to identify stars which were born in halos other than the main halo but are within the main halo at $z$=0. However, the halo tracking algorithms differ in their implementation. FIRE-2 uses the ROCKSTAR algorithm \citep{Behroozi2013,Behroozi_etal_2013b} while Auriga uses a ``Friends of Friends'' algorithm \citep{Grand2017}.Although the methods are philosophically the same, there could be implementation differences which require deeper investigation beyond the scope of this work \citep[e.g.,][]{Mansfield2024}.

\section{Stellar and dark matter profiles along the minor axis} \label{sec:prof}

\begin{table*}
\caption{Simulated Galaxy Properties}
\label{tab:table1}
\begin{tabular}{cccccccccc}
\hline\hline
Galaxy &  $\rm{M_{200c}}$ & $\rm{R_{200c}}$ & $\rm{M_{*,90}}$  & $\rm{N_{sp}}$ & $\rm{\rho_{0,H}}$ & $\rm{\alpha}$ & $\rm{\rho_{0,DM}}$ & $\rm{\alpha_{DM}}$ \\
 &  [$10^{10}~\rm{M_{\odot}}$] & [kpc] & [$10^{10}~\rm{M_{\odot}}$]  &   & [$10^{8}~\rm{M_{\odot}/kpc^3}$] &  & [$10^{8}~\rm{M_{\odot}/kpc^3}$] &  \\
\hline
m12m &  97.8 & 204 & 11 & 5& 11.1$\pm$0.00 & 3.81$\pm$0.00 &  6.90$\pm$0.00 &  1.98$\pm$0.00 \\ 
Romulus & 134 & 225 & 9.1 & 4 &  42.3$\pm$0.04 &  3.37$\pm$0.00 &  27.8$\pm$0.01 &  1.84$\pm$0.00  \\
m12b & 94.2 & 202 & 8.5  & 1 &  28.1$\pm$0.02  &  3.98$\pm$0.00  &  9.30$\pm$0.00  &  1.90$\pm$0.00 \\
m12f & 108 & 211 & 7.9 & 5 &  8.75$\pm$0.00  &  3.75$\pm$0.00  &  8.12$\pm$0.00  &  1.91$\pm$0.00 \\
Thelma & 94.1 & 200 & 7.1 &8 &  7.34$\pm$0.01  &  3.34$\pm$0.00  &  4.13$\pm$0.00  &  1.80$\pm$0.00 \\
Romeo & 93.9 & 206 & 6.6 & 14 &  12.1$\pm$0.00  &  3.83$\pm$0.00  &  20.2$\pm$0.00  &  1.92$\pm$0.00  \\
m12i & 78.3 & 190 & 6.3 & 8 &  6.74$\pm$0.00  &  3.74$\pm$0.00  &  7.84$\pm$0.00  &  1.89$\pm$0.00 \\
m12c & 91.2 & 200 & 5.8 & 3 &  2.36$\pm$0.00  &  4.23$\pm$0.03  &  6.84$\pm$0.00  &  2.03$\pm$0.00 \\
m12w & 71.8 & 188 & 5.7 & 5 &  6.14$\pm$0.00  &  3.52$\pm$0.00  &  7.10$\pm$0.00  &  1.77$\pm$0.00 \\
Remus & 85.9 & 194 & 4.6 & 10 &  8.68$\pm$0.00  &  3.46$\pm$0.00  &  11.6$\pm$0.00  &  1.89$\pm$0.00 \\
Juliet & 72.4 & 189 & 3.8 & 4 &  2.66$\pm$0.00  &  3.28$\pm$0.00  &  15.1$\pm$0.00  &  1.95$\pm$0.00 \\
Louise & 71.0 & 182 & 2.6 & 6 &  2.88$\pm$0.00  &  4.12$\pm$0.01  &  6.42$\pm$0.00  &  2.07$\pm$0.00 \\
m12r & 78.5 & 194 & 1.7& 2 &  0.471$\pm$0.00  &  4.89 $\pm$0.11  &  3.40$\pm$0.00  &  1.99$\pm$0.00  \\
 \hline
Au27 & 170 & 251 &  8.3 & 7 &  4.93$\pm$0.00  &  3.72$\pm$0.04  &  24.8$\pm$0.05  &  1.76$\pm$0.00 \\
Au21 & 142 &  237 & 7.1 & 4 &  2.48$\pm$0.00  &  2.98$\pm$0.01  &  26.3$\pm$0.17  &  1.70$\pm$0.00 \\
Au23 & 150 & 242 & 7.0 & 8 &  1.45$\pm$0.00  &  4.87$\pm$0.50  &  24.8$\pm$0.03  &  1.88$\pm$0.00 \\
Au16 & 150 & 242 & 6.5 & 8 &  0.622$\pm$0.000  &  4.65$\pm$1.26  &  33.8$\pm$0.24  &  1.89$\pm$0.00 \\
Au24 & 147 & 240 & 6.5 & 8 &  0.616$\pm$0.000  &  5.87$\pm$1.96  &  24.2$\pm$0.14  &  1.84$\pm$0.00 \\
Au6 & 101 & 212 & 5.4 & 9 &  6.00$\pm$0.01  &  4.25$\pm$0.04  &  35.0$\pm$0.13  &  1.88$\pm$0.00 \\
\hline
\end{tabular}
\tablecomments{We provide the following properties for each galaxy at $z=0$.
$\rm{M_{200c}}$ and $\rm{R_{200c}}$ are the total mass and spherical radius in which the average density is 200$\times$ the critical density of the universe. $\rm{M_{*,90}}$ is the stellar mass within a spherical radius that encloses 90\% of the stellar mass within 20 kpc \citep{Wetzel2023}. $\rm{N_{sp}}$ is the number of satellites which contribute 90\% of the accreted stellar mass \citep{Monachesi2019}. $\rm{\rho_{0,H}}$ and $\alpha$ are the stellar halo central mass density and slope, while $\rm{\rho_{0,DM}}$ and $\alpha_{DM}$ are the corresponding parameters for the dark matter component (see Appendix \ref{app:a}).   }
\end{table*}

The main focus of this work is on the cosmological predictions for minor axis stellar density profiles in the inner regions of Milky Way-mass galaxies. We define the inner region as within the NFW profile \citep{Navarro1997} scale radius of the dark matter halo. Specifically, for the FIRE-2 simulations, we use the value given from the halo finder ROCKSTAR \citep{Behroozi2013} for the host dark matter halo, rounded up to an integer. For the \textit{Latte} galaxies, the NFW scale radii range from 16-24 kpc with an average of 20 kpc. For the \textit{ELVIS} galaxies, these values range from 13-20 kpc with an average of 16 kpc. The estimate of the NFW scale radii for the Auriga galaxies comes from \citet{Callingham2020}.

We select all the stellar particles within a cylinder along the minor axis with a radius of $1/\sqrt{\pi}$ so that each cylindrical bin with height of 1 kpc has the volume of 1 $\rm{kpc^3}$. After multiplying these particles by their mass, we retrieve the total stellar mass density profile along the minor axis. We fit a simple power law and exponential disk model to this profile. Explicitly, 
\begin{equation}
    \rho_*(R) = \frac{\rho_{0,H}}{1+\left(\frac{R}{r_H}\right)^\alpha}+\rho_{0,D}e^{-R/h_D}
\end{equation} 
where $\rho_{0,H}$, and $\rho_{0,D}$ are the central halo/bulge and disk mass densities, respectively. The scale height for the disk is $h_D$, the scale radius for the halo is $r_H$ and the halo power law slope is given by $\alpha$. To avoid the impact of binning when fitting to the simulated galaxy profiles, we perform the fit to the reverse cumulative distribution function (CDF) normalized by the total mass within the NFW scale radius and $1/\sqrt{\pi}$ kpc of the minor axis. This fit provides the disk scale height, halo power law slope, halo scale radius, and the relative strength of the disk mass density compared to the halo. To derive the $\rho_{0,H}$ and $\rho_{0,D}$ we then fit to the unnormalized profile. For further details and figures demonstrating the fits, we refer the reader to Appendix \ref{app:a}.

In addition to quantifying the density profile, we are interested in determining the radius at which halo substructures become significant. To do this we measure the density profile along different lines and compare to the profile along the minor (z-)axis. Specifically, we rotate the the simulated galaxy about the x-axis by a random angle, $\phi$, between 0 and $\pi/4$ radians, in order to avoid the galactic disk. We then rotate the galaxy again but this time about the z-axis by a random angle, $\theta$, between 0 and 2$\pi$ radians. Each time we measure the mass density profile along the new z-axis using the same cylinder as before. We perform 100 unique combinations of rotations to obtain 100 different estimates of the mass density profile. We then compute the fractional difference between this and the original, non-rotated z-axis profile (see Figure \ref{fig:si}-\ref{fig:corr} in Appendix \ref{app:a}). 

In order to understand how the stellar component relates to its dark matter counterpart, we also fit the dark matter halo distribution using the same technique. The only difference is that we do not include a exponential disk component in the functional form. Therefore, the functional form of the dark matter mass density distribution is the following power law:
\begin{equation}
    \rho_{DM}(R) = \frac{\rho_{0,DM}}{1+\left(\frac{R}{r_{DM}}\right)^{\alpha_{DM}}}
\end{equation} 
where $\rho_{0,DM}$ is the dark matter mass density at $R$ = 0, $r_{DM}$ is the scale radius and $\alpha_{DM}$ is the power law slope. We follow the same procedure as the stellar component for fitting the functional form as well as the angle dependence. We find that the fractional difference between the dark matter density profiles at random angles is significantly smaller at all radii than that of the stellar profiles, indicating that the stellar particles have more significant substructure than the dark matter.

\section{Inner Minor Axis Stellar Density Profiles are Weakly Correlated with Accretion History} \label{sec:not_accr}

\begin{figure*}
    \centering
    \includegraphics[width=\linewidth]{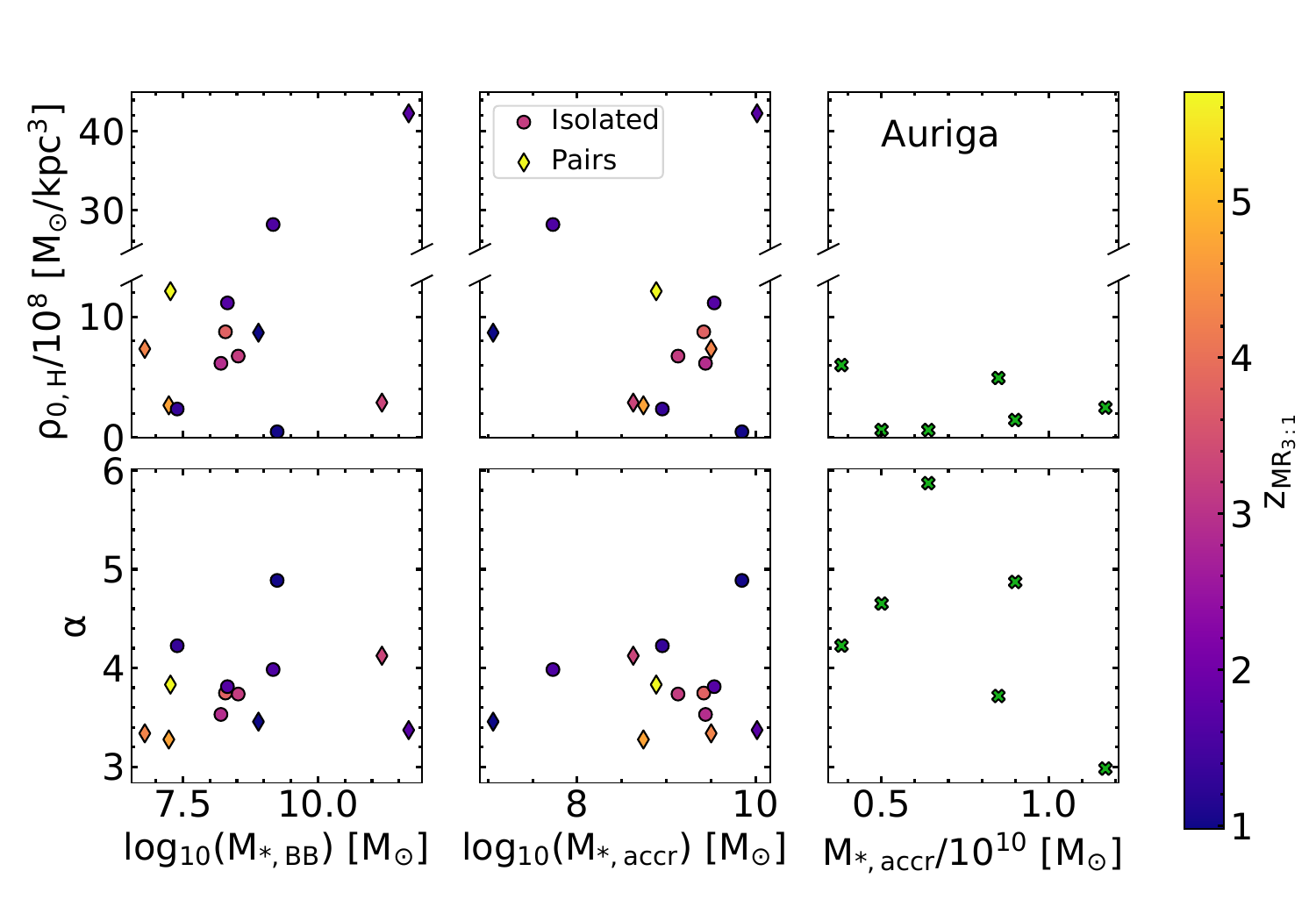}
    \caption{ How the minor axis stellar density profile parameters relate to metrics of the accreted mass in the simulated galaxies. Specifically, the top row shows the stellar density amplitude ($\rho_{0,H}$) on the y-axis, while the bottom row shows the stellar density slope ($\alpha$). The left and middle panels show results for the FIRE-2 simulations, colored by $z_{MR_{3:1}}$. The left panel has the stellar mass of the building blocks (i.e., systems accreted before $z_{MR_{3:1}}$), while the middle panel shows the total stellar mass accreted after $z_{MR_{3:1}}$. The right panels shows results for the Auriga galaxies in green. }
    \label{fig:ma}
\end{figure*}

\begin{figure*}
    \centering
    \includegraphics[width=\linewidth]{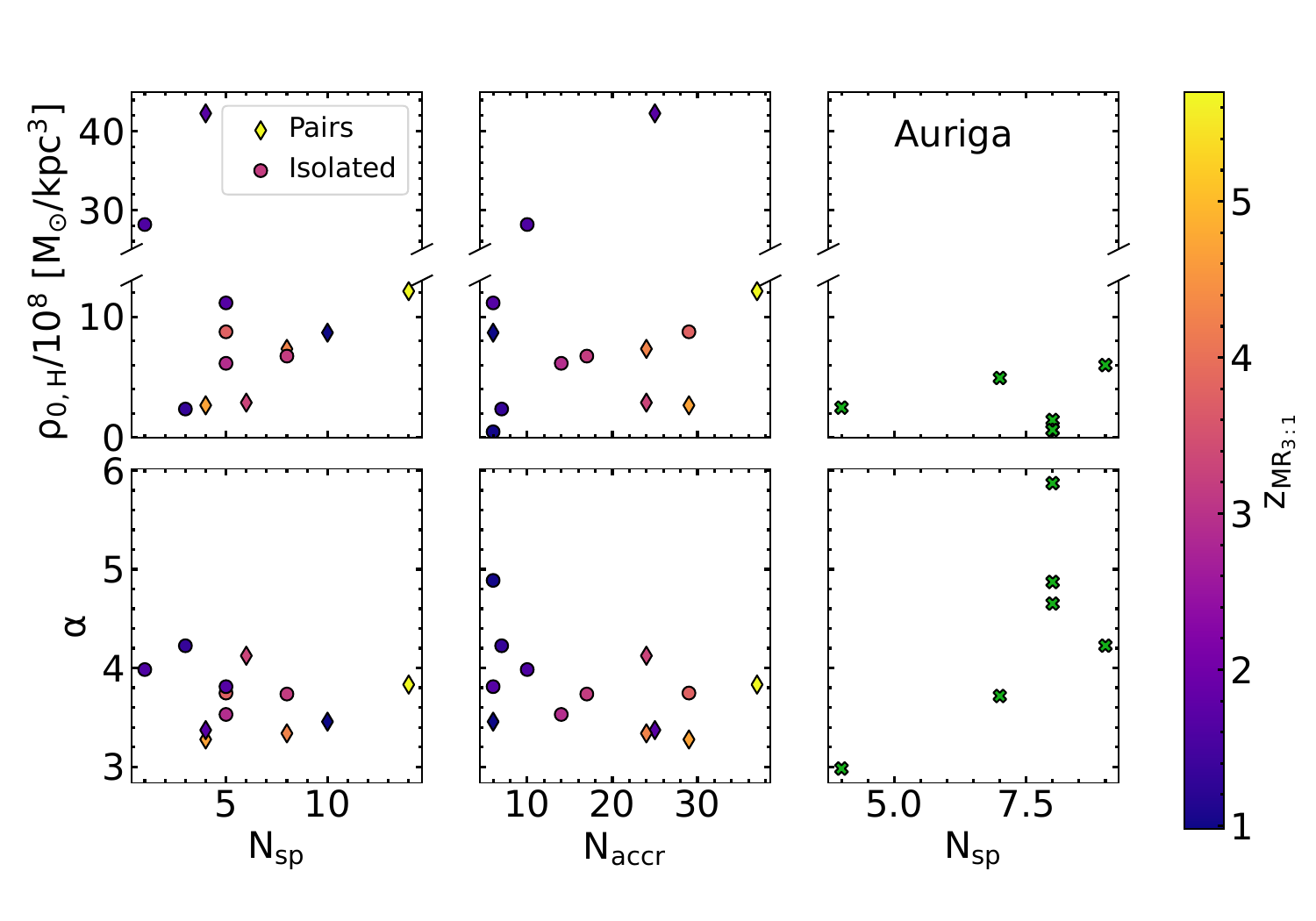}
    \caption{ Similar to Figure \ref{fig:ma}, but the horizontal axes now show the number of significant progenitors (left and right panels) or accreted systems (middle panel). As with Figure \ref{fig:ma}, there are not strong correlations between the minor axis stellar density profiles and the accretion history metrics.  }
    \label{fig:na}
\end{figure*}

It is thought that the halo stellar density profile traces the merger history of a galaxy \citep{Bullock2005,Cooper2010,Deason2013,Monachesi2019,D'Souza2018}, but is is unclear how this might extend to the innermost regions of the Galaxy. Furthermore, mergers can dynamically disturb galaxies and build up a centrally concentrated in-situ halo component \citep{Grand2020,Dillamore2022}. The minor-axis profiles of the inner regions of galaxies may be dominated by an in-situ halo component or bulge. We note, however, that surface brightness profiles of the simulated galaxies used in this work indicate that none of them have a classical bulge \citep{Sanderson2018,Gargiulo2019}. Furthermore, we find that the minor-axis stellar density profiles out to 20 kpc are best fit by only an exponential disk with scale height $<$0.5 kpc and a single power-law model.

In general, galaxy formation simulations can form galactic bulges through merger events and secular evolution processes \citep{Somerville2015}. Therefore, it is uncertain whether bulge mass traces the galaxy's merger history.  Using the Auriga galaxies, \citet{Gargiulo2019}, did not find a significant correlation between the bulge mass and the accreted stellar halo mass. Observationally, this is difficult to test given that a galaxy's accretion history is not directly observable. Nonetheless, using 18 MW-mass galaxies, \citet{Bell2017} did not find strong correlations between the stellar halo mass and the bulge mass, indicating the bulges may gain mass through secular processes. In this section, we directly investigate whether there is a correlation between the inner stellar density profile along the minor axis and the accretion history of the FIRE-2 and Auriga galaxies.  

In the inner regions of galaxies, it is difficult to distinguish between the bulge and stellar halo given that they spatially and kinematically overlap. Furthermore, there is no universally accepted definition to disentangle these components, even in simulations with complete stellar particle data. Therefore, in this work we investigate the total stellar density distribution along the minor-axis, facilitating straightforward comparisons to observations in future work.

We compare measured properties of the $z=0$ minor axis stellar density profiles of Milky Way-mass galaxies (see Section \ref{sec:prof}) with their accretion histories. In order to quantify and interpret strengths of correlations in this work, we use the Spearman correlation coefficient which measures the monotonicity of the relationship between two variables \citep{Spearman,dancey2004statistics}. Therefore, the Spearman correlation coefficient will be close to 1 if the data have similar rank in the two variables, i.e., the data point with the largest $x$-value also has the largest $y$-value and so on. Furthermore, the Spearman correlation coefficient will be close to -1 if the data have almost opposite rank. Generally, a correlation is considered weak for Spearman coefficients with an absolute value <0.40. A moderate correlation has an absolute Spearman coefficient between 0.40 and 0.60, while a strong correlation has >0.60. The uncertainties on the Spearman correlation coefficients are calculated by bootstrapping the uncertainties on the density profile fit parameters ($\alpha$,$\rho_{0,H}$,$\alpha_{DM}$,$\rho_{0,DM}$).

In Figure \ref{fig:ma}, we show how the amount of accreted stellar mass in each galaxy relates to $\rm{\alpha}$, the power law slope (bottom panels), and $\rm{\rho_{0,H}}$, the central stellar mass density for the power-law component (top panels). The left and middle panels show results for the FIRE-2 galaxies while the right panels show results for the Auriga galaxies (green Xs). The FIRE-2 pair galaxies are shown as diamonds, while the isolated galaxies are circles. The FIRE-2 results are colored by the redshift when the proto-Milky Way emerges, $\rm{z_{MR_{3:1}}}$, defined in \citet{Horta2024} as the redshift when the main galaxy halo becomes 3 times more massive than the next most massive luminous halo. For the Auriga galaxies the horizontal axis is the reported accreted stellar halo mass from \citet{Monachesi2019}. For the FIRE-2 galaxies we use two different metrics for the accreted stellar mass. The left panel horizontal axis is the $\rm{log_{10}}$ of the stellar mass of the simulated galaxy's building blocks from \citet{Horta2024}. Specifically, the building blocks are defined as luminous halos which merge with the main branch before $\rm{z_{MR_{3:1}}}$. In the middle panel, the horizontal axis is the $\rm{log_{10}}$ of the phase-mixed stellar mass accreted after $\rm{z_{MR_{3:1}}}$ from Kundu et al. (in prep).

It is interesting to note that although the Auriga central stellar densities overlap with the FIRE-2 results, the range is smaller: all Auriga halos have $\rm{\rho_{0,H}}<10^9$ \densu\ while almost a third (4/15) FIRE-2 galaxies have power law components with central densities larger than that. The opposite is true for the power law slopes, $\alpha$, in that the Auriga galaxies have larger scatter than FIRE-2. However, we note that these differences are likely due to fit degeneracies. There is an especially strong degeneracy between $\rm{\rho_{0,H}}$ and $\rm{r_H}$ in that smaller $\rm{r_H}$ indicates a more cuspy density profile with a higher central density. We note that the FIRE-2 galaxies with larger $\rm{\rho_{0,H}}$ all have $\rm{r_H}$ that are smaller than the Auriga galaxies (see Appendix \ref{app:a}).

Both Auriga and FIRE-2 show weak correlations between the various measures of accreted stellar mass and the minor axis stellar density profile.  For the FIRE-2 galaxies, the strongest correlation is between the power law slope and the total stellar mass of the building blocks. The Spearman correlation coefficient for these two measures is 0.36$\pm$0.05, which signifies a weak relationship. Similarly, for the Auriga galaxies the power law slope is most correlated with the accreted stellar mass, with a Spearman correlation coefficient of -0.40$\pm$0.20. We note that the uncertainty on the correlation coefficient is large here because of the significant uncertainties on the power law slopes for the Auriga galaxies (see Table \ref{tab:table1}).  In conclusion, we do not find strong correlations between the accreted stellar mass and the amplitude or slope of the minor axis stellar density profile. 

Figure \ref{fig:na} is the same as Figure \ref{fig:ma} except for the horizontal axes. In Figure \ref{fig:na}, the horizontal axis for the left and right panels is the number of ``significant progenitors" from \citet{Monachesi2019}, defined as the number of satellites that together contribute 90\% of the total accreted stellar mass. The middle panel horizontal axis is the number of phase- mixed accreted systems from Kundu et al. (in prep). 

The Auriga galaxies have a positive correlation between the number of significant progenitors and the stellar density profile slope with a Spearman correlation coefficient of 0.59$\pm$0.11. This is classified as a moderate relationship. However, a negative correlation is reported in \citet{Monachesi2019} with the 28 Milky Way-mass Auriga galaxies run at lower resolution, although the correlation is weak. We also see a negative correlation between the stellar density profile slope and number of accreted systems with the FIRE-2 galaxies, but the correlation is weak with a Spearman correlation coefficient of -0.39$\pm$0.04. In general, we find inconclusive evidence that the number of merger events impacts the minor axis stellar density profile for the inner regions of these simulated galaxies.

\section{The Stars Follow the Dark Matter} \label{sec:dm}

\begin{figure*}
    \centering
    \includegraphics[width=\linewidth]{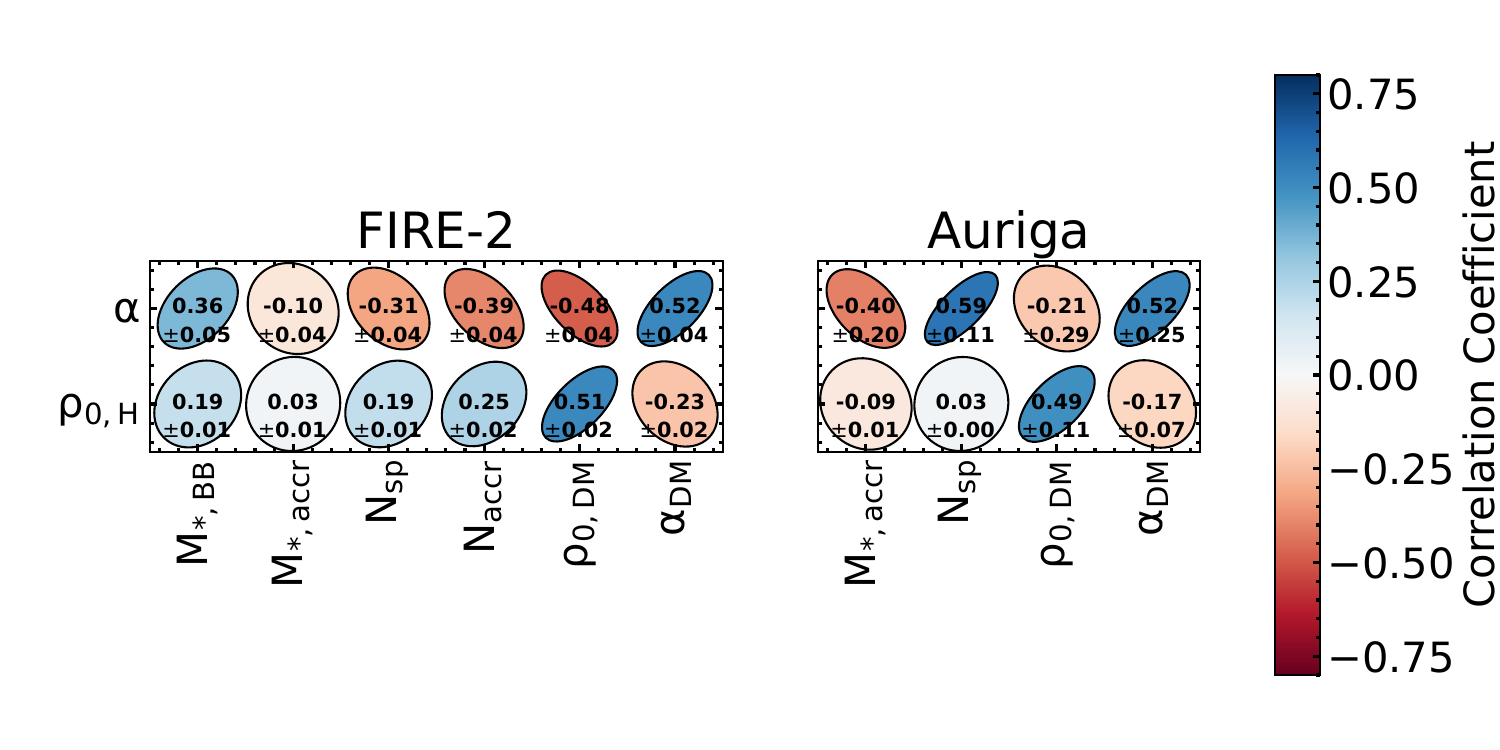}
    \caption{The Spearman correlation coefficients for the minor axis stellar density profile parameters correlated with metrics of the accretion history and the minor axis dark matter density profile. Specifically, we show an ellipse whose shape, color and orientation demonstrate the coefficient's value. The value of each correlation coefficient is printed on the ellipse with the corresponding uncertainty printed below. Large positive (negative) correlation coefficients are shown with bluer (redder) ellipses that are more elongated to the top-right (bottom-right). The top row shows the correlations with the minor axis stellar density slope ($\alpha$), while the bottom row shows the correlations with the amplitude of the profile ($\rho_{0,H}$). The left panel shows results for the FIRE-2 simulations with stellar mass of the building blocks ($M_{*,BB}$), total stellar mass of accreted systems ($M_{*,accr}$), number of significant progenitors ($N_{sp}$), number of accreted systems ($N_{accr}$), minor axis dark matter density profile amplitude ($\rho_{0,DM}$) and slope ($\alpha_{DM}$), from left to right. The right panel shows the results for the Auriga galaxies with the stellar mass of accreted systems ($M_{*,accr}$), number of significant progenitors ($N_{sp}$), minor axis dark matter density profile amplitude ($\rho_{0,DM}$) and slope ($\alpha_{DM}$), from left to right.   }
    \label{fig:corr}
\end{figure*}

\begin{figure}
    \centering
    \includegraphics[width=\linewidth]{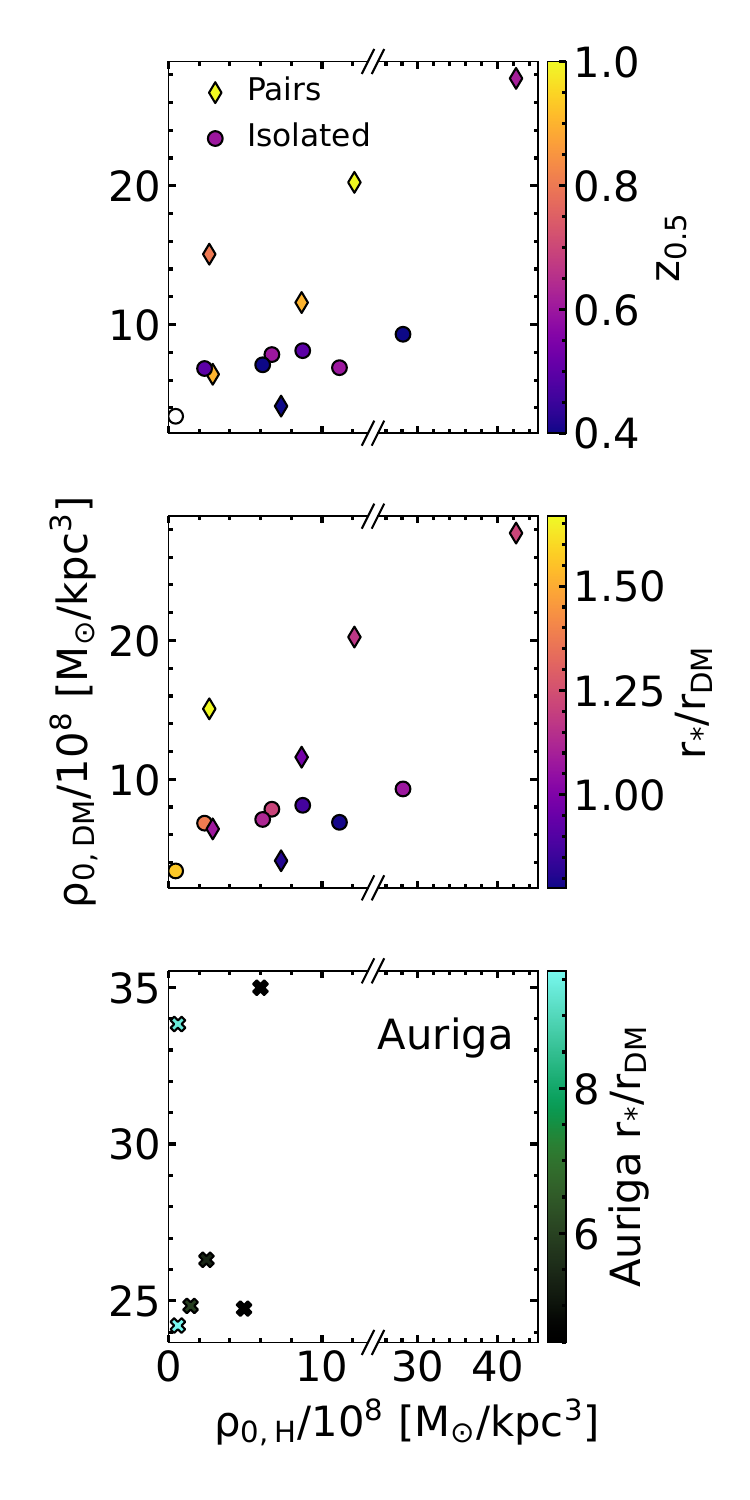}
    \caption{ The relationship between the minor axis stellar density profile amplitude and the dark matter counter-part. The results for FIRE-2 simulations are shown in the top two panels, while results for the Auriga galaxies are in the bottom plot. In the top plot, the points are colored by the formation redshift ($z_{0.5}$). For the bottom two plots the points are colored by the ratio of the scale radius for the stellar and dark matter profiles. The Auriga galaxies use a different colormap to emphasize the difference in range of the scale radius ratios.   }
    \label{fig:ns}
\end{figure}

\begin{figure}
    \centering
    \includegraphics[width=\linewidth]{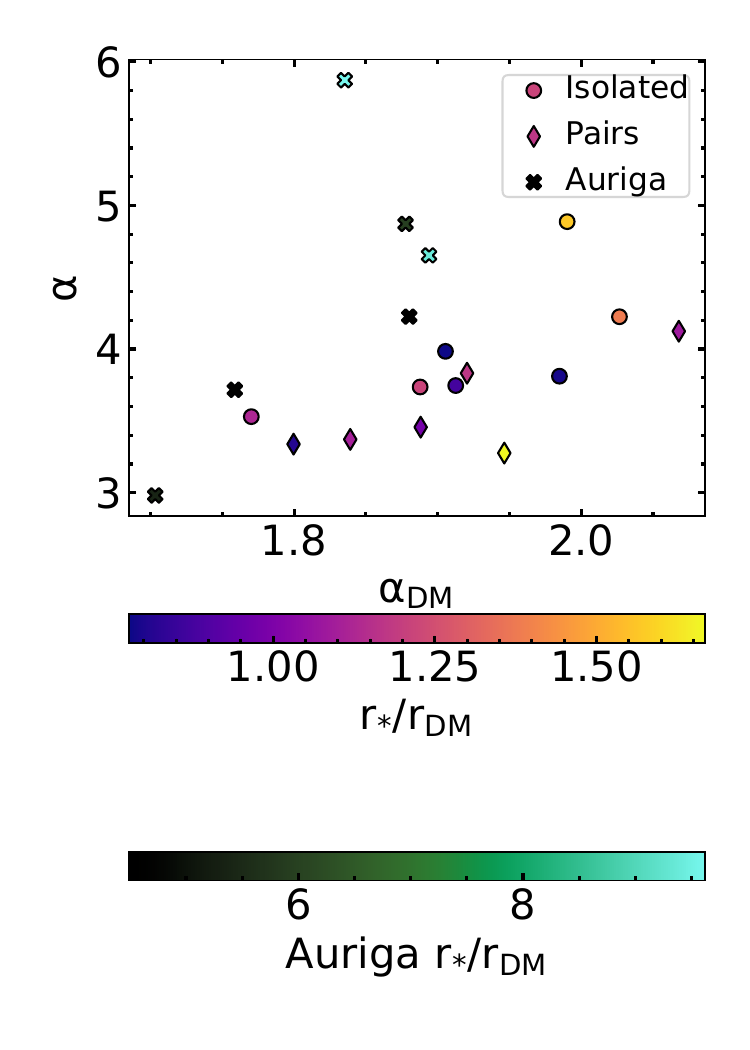}
    \caption{The relationship between the slope of the minor axis stellar density profile with the dark matter counterpart. The FIRE-2 galaxies are colored by the ratio of the scale radii of the stellar and dark matter profiles with isolated galaxies as circles and pairs as diamonds. The results for the Auriga galaxies are shown as Xs also colored by the ratio of the scale radii, but we use a different colormap to emphasize the difference in range compared to the FIRE-2 galaxies.    }
    \label{fig:slope}
\end{figure}

\begin{figure*}
    \centering
    \includegraphics[width=\linewidth]{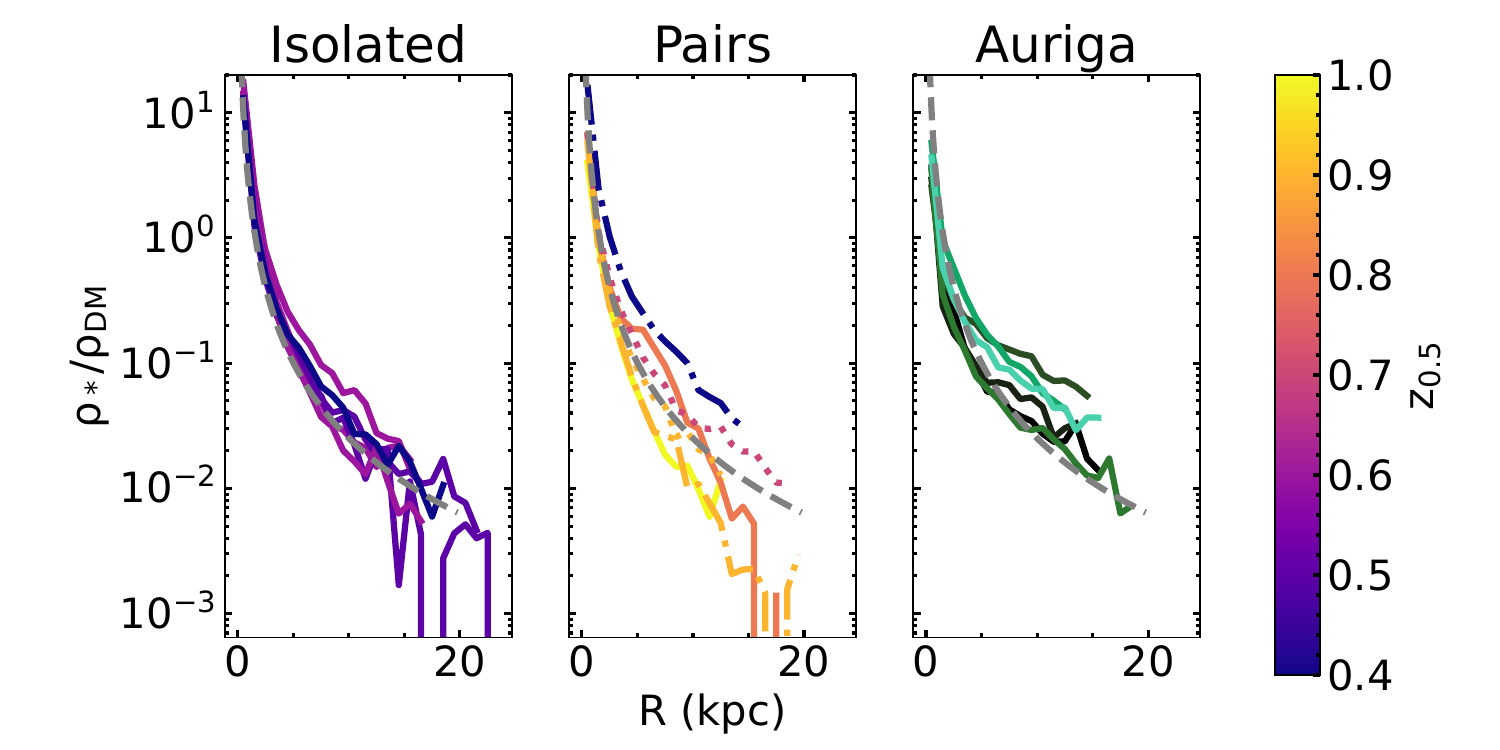}
    \caption{ The ratio of total stellar density to dark matter density in the simulated galaxies as a function of increasing distance in the direction that is perpendicular to the galaxy's disk plane. The left panel shows results for the isolated FIRE-2 galaxies while the middle panels shows results for the pairs. In both the left and middle panels, the FIRE-2 galaxies are colored by their formation redshift ($z_{0.5}$). The right panel shows results for the Auriga galaxies in random shades of green. The grey dashed line shows the best fit to the FIRE-2 galaxies which is simply $\rho_*/\rho_{DM} = 0.4/R^2 $.  } 
    \label{fig:ratio}
\end{figure*}

In this section, we compare the measured minor axis stellar density profiles with the corresponding dark matter density profiles. In general, we find that the strength and slope of the stellar profile is highly correlated with the dark matter profile. 

In Figure \ref{fig:corr}, we show a graphical representation of the correlations between the minor axis stellar and dark matter density profile parameters along with parameters describing the accretion history of the galaxy. Specifically the Spearman correlation coefficients are represented as ellipses with ellipticity, orientation and color based on the their value. Strong positive correlations are shown with narrow ellipses that point up to the right in dark blue while strong negative correlations point down to the right in dark red. Weak correlations with coefficients close to zero are shown as white circles. Results for how the amplitude and slope of the FIRE-2 galaxies' minor axis stellar density profiles relate to the stellar mass of the building blocks, accreted stellar mass, number of significant progenitors, and number of accreted systems, along with the minor axis dark matter density profile's amplitude and slope are shown on the left in order from left to right. Results for the correlations of the amplitude and slope of Auriga's minor axis stellar density profiles with the accreted stellar mass, number of significant progenitors, minor axis dark matter density profile amplitude and slope are shown on the right, also in order from left to right. 

We show the correlations for the stellar mass of the building blocks, the accreted stellar mass, the number of accreted systems and the number of significant progenitors in Figure \ref{fig:corr} in order to compare with the dark matter correlations, although they have already been discussed in Section \ref{sec:not_accr} (see Figures \ref{fig:ma} and \ref{fig:na}). 

For the FIRE-2 galaxies the strongest correlations are between the corresponding minor-axis stellar and dark matter density profile parameters. Specifically, the stellar and dark matter density amplitudes are strongly positively correlated, along with the slopes. On the other hand the stellar density amplitude is negatively correlated with the dark matter slope and the dark matter density amplitude is negatively correlated with the stellar slope. This is likely due to the combination of the degeneracies in the fit of the density amplitudes and slopes for both components and the strong positive correlations between the components' profiles. When comparing across the stars and dark matter, the strength of the positive correlations between the components' amplitudes and slopes are 0.51$\pm$0.02 and 0.52$\pm$0.04, respectively, which are considered moderate relationships. The negative correlation coefficient between the stellar amplitude and dark matter slope is -0.23$\pm$0.02, while the dark matter amplitude and stellar slope have a correlation coefficient of -0.48$\pm$0.04. 

For the Auriga galaxies, the correlation coefficient of the minor axis stellar density profiles with their corresponding dark matter density profiles have larger uncertainties  than the FIRE-2 galaxies. Specifically, the correlations between the amplitude and slope are both 0.49$\pm$0.11 and 0.52$\pm$0.25, which are both considered moderate, but with larger uncertainties than the FIRE-2 correlations of 0.52$\pm$0.04 and 0.51$\pm$0.02. This is likely because the functional form was optimized for the FIRE-2 galaxies and therefore the Auriga galaxies generally have larger uncertainties on the fit parameters. Specifically, the average $log_{10}$ root-mean-squared error (RMSE) is 7.38 for Auriga galaxies while it is only 7.16 for FIRE-2 galaxies. We further discuss the relative goodness-of-fits in Appendix \ref{app:a}.

In Figure \ref{fig:ns}, we show the relationship between the minor axis stellar density profile amplitude with the corresponding dark matter density amplitude for the FIRE-2 Milky Way-mass galaxies (top two panels) and the Auriga Milky Way-mass galaxies (bottom panel). In the top panels, the FIRE-2 galaxy pairs are shown as diamonds while the isolated galaxies are circles. In the topmost panel, the points are colored by their formation time. Specifically, the color corresponds to the redshift at which the galaxy has reached 50\% of its z=0 mass ($\rm{z_{0.5}}$). The middle panel is the same as the top panel, except here the points are colored by the ratio of the stellar to dark matter scale radii. The bottom panel shows the same quantities for the Auriga galaxies, but  the color bar scale is quite different than the middle panel's. 

Not only do the ratios of the scale radii for the stellar and dark matter density profiles have larger values, but the Auriga halos also have larger dark matter density amplitudes than the FIRE-2 galaxies on average. For example, only 2 out of 15 FIRE-2 galaxies have $\rm{\rho_{0,DM}}$ > 20 $\times10^8$ \densu\ while the least dense Auriga halo has $\rm{\rho_{0,DM}}$ = 24 $\times10^8$ \densu . This is consistent with the fact that the FIRE-2 and Auriga galaxies have a similar range of $M_{200}\approx$ 1-2 $\times10^{12}~M_{\odot}$ but FIRE-2 has $R_{200m}\approx$320-410 kpc while Auriga has $R_{200}\approx$210-260 kpc. The virial radii $R_{200}$ are calculated differently in that Auriga uses the radius within which the density is 200$\times$ the critical density for closure, while FIRE-2 uses the radius within which the density is 200$\times$ the mean density of the Universe. Either way, the Auriga galaxies have similar mass inside a smaller radius than the FIRE-2 galaxies, and therefore higher mass density. The Auriga dark matter density profiles also have smaller scale radii than the FIRE-2 dark matter profiles, while the stellar profiles have larger scale radii on average. This is likely due to differences in feedback mechanisms between the simulations, which are discussed further in Section \ref{sec:diffs}.

In the top panel of Figure \ref{fig:ns}, the FIRE-2 galaxy pairs with early formation times are outliers in the relationship between the stellar and dark matter density amplitudes. One possible reason for this is simply that these structures formed earlier when the universe had an overall lower stellar-to-dark-matter mass ratio. In the middle panel, we demonstrate how the degeneracy of the fits impacts the spread in the relationship between the stellar and dark matter density amplitudes for the FIRE-2 galaxies. On the other hand, the Auriga galaxies do not show a strong correlation between the stellar and dark matter density amplitudes. However, given the consistency of the ratio of stellar-to-dark-matter mass seen in Figure \ref{fig:ratio}, the correlation is likely masked by degeneracies in the fit parameters (see Appendix \ref{app:a}).

The relationship between the minor axis stellar density power law slopes and the corresponding dark matter slopes are shown in Figure \ref{fig:slope}. The Auriga galaxies are shown as Xs while the FIRE-2 isolated galaxies are shown as circles and the pairs as diamonds. The points are colored by the ratio of their stellar and dark matter scale radii, but as in Figure \ref{fig:ns}, the Auriga galaxies have a different color scale than the FIRE-2 galaxies. 

The Auriga and FIRE-2 galaxies both show a strong correlation between their stellar and dark matter density slopes. In general, the Auriga galaxies have steeper stellar slope for a given dark matter slope.  Similar to the stellar density amplitudes, the largest outliers in the relationship between the stellar and dark matter density slopes for the FIRE-2 galaxies are also outliers in the ratio of the corresponding scale radii. This indicates much of the spread is likely due to degeneracies in the fit parameters.

Figure \ref{fig:ratio} shows the ratio of total stellar density to dark matter density for the simulated galaxies as a function of distance from the galactic center. Specifically, we calculate these ratios similar to the density profile fits in Section \ref{sec:prof} in that we only use particles within a cylinder along the minor axis with radius=$1/\sqrt{\pi}$. We utilize all the particles within this cylinder, including the disk particles. In the left panel we show results for the isolated FIRE-2 galaxies, while the middle panel shows the pairs. The lines in these panels are colored by the redshift at which the galaxy has reached 50\% of its $z=0$ mass. In the middle panel, we also use different line styles for sets of pairs. Specifically, Romeo and Juliet are shown as solid lines, while Romulus and Remus are in dotted lines and Thelma and Louise in dot-dashed lines. The right panel shows results for the Auriga galaxies colored in random shades of green. Each panel also has a grey dashed line which is simply the function $\rm{\rho_*/\rho_{DM}=0.4/R^2}$ which is the best fit to the results for the FIRE-2 pair galaxies (middle panel).

The ratio of stellar to dark matter density along the minor axis in these simulated galaxies follows a $\rm{r^{-2}}$ profile. While each simulation suite generally follows this relation, the FIRE-2 and Auriga galaxies show slightly different behavior. The later forming (lower $\rm{z_{0.5}}$) FIRE-2 pairs have higher $\rho_*/\rho_{DM}$ at $R>10~kpc$ compared to the earlier forming galaxies. The Auriga galaxies generally have higher stellar to dark matter ratios similar to the later forming FIRE-2 galaxy pairs. It is possible that this is because they also form later, when the universe overall has a higher stellar-to-dark-matter density, but we do not have formation redshifts for these galaxies to confirm.

\section{Discussion and Conclusions} \label{sec:conclu}

In this work, we parameterized the minor axis stellar and dark matter density profiles of Milky Way-mass FIRE-2 and Auriga simulated galaxies. With this parameterization, we compare the stellar profiles to the accretion histories and find weak to no correlation. Instead, we find strong correlations between the distribution of stars and dark matter. Specifically, we find that the ratio of stellar to dark matter density consistently falls off as $\rm{r^{-2}}$ along the minor axis across the simulation suites.

Whether the consistency of the stellar-to-dark-matter profile is a natural consequence of galaxy formation and \lcdm, or if its merely a coincidence in these simulations is unknown. Consistent with the profiles measured in this work, dark matter halos are generally thought to have pseudo-isothermal profiles with $\rm{r^{-2}}$ within their NFW scale radii \citep{Navarro1997}. Predictions for the minor axis stellar density profiles in this range are fewer and vary more widely, but generally cluster around $\rm{r^{-4}}$ \citep{Monachesi2019,Deason2013,Font2020,Cooper2010,Amorisco2017}. In this work, we find that degeneracies in the fit parameters can cause more spread in the measured slopes (see Figure \ref{fig:slope}), but the profile of the ratio of stellar-to-dark-matter is consistent (see Figure \ref{fig:ratio}). 

The idea that the stellar and dark matter profiles might be correlated is not new. Baryonic feedback and adiabatic contraction are known to impact the density profiles of dark matter halos \citep{Blumenthal1986,DiCintio2013,Pontzen2014,Chan2015,Lazar2020,Callingham2020}. However, the most interesting result of our work is that both FIRE-2 and Auriga have a stellar to dark matter ratio that falls  off as $\rm{r^{-2}}$ along the minor axis \textit{despite} the differences in their baryonic physics (see discussion in Section \ref{sec:diffs}). 

We have intentionally investigated the  minor axis stellar density profiles in a region where galactic component classification is difficult. Specifically, the radial range we study is where the disk, bulge and inner halo overlap. Although it is clear that the exponential component of the stellar density profile is the galactic disk, we do not attempt to classify the power-law component. It could be classified as either bulge or inner halo. Classical bulges have long been observed to have $\rm{r^{-4}}$ profiles, similar to elliptical galaxies \citep{deVaucouleurs1948,Hernquist1990}. However, estimates for the S\'ersic indices of the FIRE-2 and Auriga galaxies have $n<2$, inconsistent with classical bulges \citep{Sanderson2018,Gargiulo2019}.  Simulations show that that $\rm{r^{-4}}$ stellar profiles can be reproduced with dissipationless collapse \citep{vanAlbada1982} and also with dissipational mergers \citep[e.g.,][]{Hernquist1992,Barnes1992}. Halo stellar density profiles are less often studied, given their low surface brightness. However, the studies that exist have found profiles with slopes that vary around $\rm{r^{-4}}$ for nearby Milky Way-mass disk galaxies \citep{Harmsen2017}. In Section \ref{sec:mw}, we discuss comparisons to the Milky Way. 

\subsection{Differences between FIRE-2 and Auriga} \label{sec:diffs}

In this work, we note a number of differences between the Auriga and FIRE-2 minor axis stellar and dark matter profiles. While we do not focus on the galactic disks in this work, we find that Auriga's disks have 10-100$\times$ higher stellar mass densities than the FIRE-2 disks (see Appendix \ref{app:a}). Relatedly, the Auriga galaxies are known to host stronger bars than the FIRE-2 galaxies \citep{Ansar2023,Fragkoudi2024}. 

Another large difference between the FIRE-2 and Auriga galaxies are the dark matter central densities ($\rm{\rho_{0,DM}}$) and scale radii ($r_\textrm{DM}$). Although they have never been directly compared, it is known that the Auriga galaxies have significant baryon contraction \citep{Callingham2020}, while the FIRE-2 galaxies form dark matter cores \citep{Lazar2020}. This is likely due to differences in the implementation of feedback, which has been shown to significantly impact the distribution of mass in the center of galaxies \citep{Pontzen2014,Lazar2020}.

The formation of dark matter halo cores in cosmological simulations is dependent on the the star formation prescriptions as they relate to the supernova rates. Specifically, the choice of gas density threshold for star formation has been shown to impact the dark matter distribution in galaxies \citep{Dutton2019,Benitez-LLambay2019}. Given that the FIRE-2 galaxies have a higher gas density threshold for star formation ($n_{SF}>1000~\rm{cm^{-3}}$) than the Auriga simulations ($n_{SF}>0.13~\rm{cm^{-3}}$), it is not surprising that they have dark matter cores ($r_{DM} \sim$ 1~kpc) and while the Auriga simulations do not ($r_{DM} \sim$ 0.3~kpc).

As discussed in Section \ref{sec:asims}, the supernova prescriptions are also different between the simulations in that FIRE-2 have rates that evolve with the star particles \citep{Hopkins2018}, while the Auriga simulations have instantaneous supernova feedback at the formation time of the star particle. FIRE-2 also implements radiative feedback from massive stars, which regulates star formation \citep{Hopkins2014,Orr2018,Hopkins2020}. Auriga does not include early stellar feedback, but does include AGN feedback, which is missing from the FIRE-2 physics model.

In total, there are many differences between the implementations of the simulations which could explain the small variations in the stellar and dark matter minor axis profiles. Despite these differences, however, the most interesting result is the \textit{remarkable similarity} of their stellar-to-dark-matter ratio profiles, which consistently fall off as $\rm{r^{-2}}$ for both simulation suites.

\subsection{Comparison to Milky Way} \label{sec:mw}

As discussed in the introduction (Section \ref{sec:intro}), there are a number of estimates of the MW halo stellar density profile. Each of these measurements have their own selection functions, biases and radial ranges. Therefore, it is difficult to compare our results without a thorough investigation of the impacts of each selection function and method. We plan to perform a detailed comparison with MW data and study the impact of selection functions in future work. Here, we briefly summarize the MW results which are most suitable for comparison with the simulated galaxies.

Generally, the simulated galaxies have steeper stellar density slopes than what is measured in the MW. Between $\approx$5-20 kpc estimates for the MW's halo stellar density profile slope range from -1 to -3 \citep{Sesar2013,Faccioli2014,Sesar2010,Han2022}. There are few measurements within 5 kpc of the Galactic center given the high levels of dust extinction and crowding. However, using RR Lyrae, the stellar density profile slope was estimated as -3 in the radial range of 0.2-2.8 kpc from the Galactic center \citep{Pietrukowicz2015}.  Furthermore, using orbit integration, \citet{Yang2022} found that the halo stellar density profile flattens in the inner regions with the slope increasing to -1.5. Although these works provide estimates in this range, they are based only on a small fraction of the stellar population, with complex selection functions. In order to perform a fair comparison, the MW minor axis profile should model the Galactic disk to minimize the selection bias. We plan to do this in future work using data with a well-modeled selection function.

The MW's dark matter density profile is not well constrained because methods rely on the poorly constrained baryonic matter distribution. Recent measurements of the Milky Way's circular velocity curve have provided estimates of the dark matter profile which give results consistent with the profiles of FIRE-2 galaxies \citep{Ou2024}. However, we note that these results give a larger core and more concentrated dark matter halo than what is typically seen in Milky Way-mass cosmological simulations \citep{Lazar2020}.

The ratio of stellar-to-dark matter mass profile has not been directly measured in the Milky Way. However, combining results from the individually measured profiles indicate a ratio profile that is roughly constant or linear, given that the stellar profile is less steep than in the simulated galaxies. If this discrepancy is borne out in future work, this could indicate a need for adjustments in the physics models used in FIRE-2 and Auriga simulations, likely either in the stellar feedback prescription, or the dark matter model.

\subsection{Summary of Conclusions}

\lcdm\ cosmology predicts that hierarchical growth dominates early galaxy formation and evolution. The minor axis stellar density profile of galaxies is of special interest as it is thought to trace the galaxy's accretion history. Whether classified as halo or bulge, this stellar component is thought to build mass through mergers and accretion events. 

In this work, we have investigated the minor axis stellar and dark matter density profiles of Milky Way-mass galaxies  in the cosmological zoom-in simulation suites of FIRE-2 and Auriga. Using only data within their respective NFW scale radiii, we quantify the profiles with a simple power law parameterization, including a exponential disk component for the stars. With the aim of understanding the physical mechanism that shapes the minor axis stellar density profile, we compare the parameters with other properties of the simulated galaxies. In total, we find:

\begin{itemize}
    \item The amplitude and slope of the minor axis power law stellar density profiles do not relate to the galaxy's accretion history.
    \item Instead, the stellar profiles are tightly correlated with the corresponding dark matter profile. Therefore, we find the dark matter potential is the dominate mechanism determining the amplitude and slope of minor axis power law stellar profile.
    \item The ratio of stellar-to-dark-matter mass decreases as $1/r^2$ along the minor axis for all simulated galaxies.
\end{itemize}

As the ratio of light to dark matter is a fundamental observation, it is crucial to understand the interplay of stellar and dark matter mass in galaxies. In future work, we plan to build on this work by investigating the comparison between the MW and simulations more deeply. This includes a new measurement of the minor axis stellar density profile using \textit{Gaia} data, and an investigation of the Milky Way's ratio of stellar-to-dark matter mass profile using stellar kinematics. Finally, we will also look to alternative dark matter models including the FIRE-2 self-interacting dark matter simulations \citep{Sameie2021,Vargya2022} to further study the connection between stellar and dark matter mass.

\software{Astropy \citep{astropy:2013,astropy:2018},
Matplotlib \citep{matplotlib},
IPython \citep{ipython},
Numpy \citep{numpy}, 
Scipy \citep{scipy},
GizmoAnalysis \citep{GizmoAnalysis}
HaloAnalysis \citep{haloanalysis,Wetzel2016}
}    

\section*{Acknowledgements}

\begin{acknowledgments} This material is based upon work supported by the National Science Foundation under Award No. 2303831. We have used simulations from the Auriga Project public data release \citep{Grand2024} available at \url{https://wwwmpa.mpa-garching.mpg.de/auriga/data}.

\end{acknowledgments}

\appendix 

\section{Density Profile Fits} \label{app:a}

\begin{figure}
    \centering
    \includegraphics[width=\linewidth]{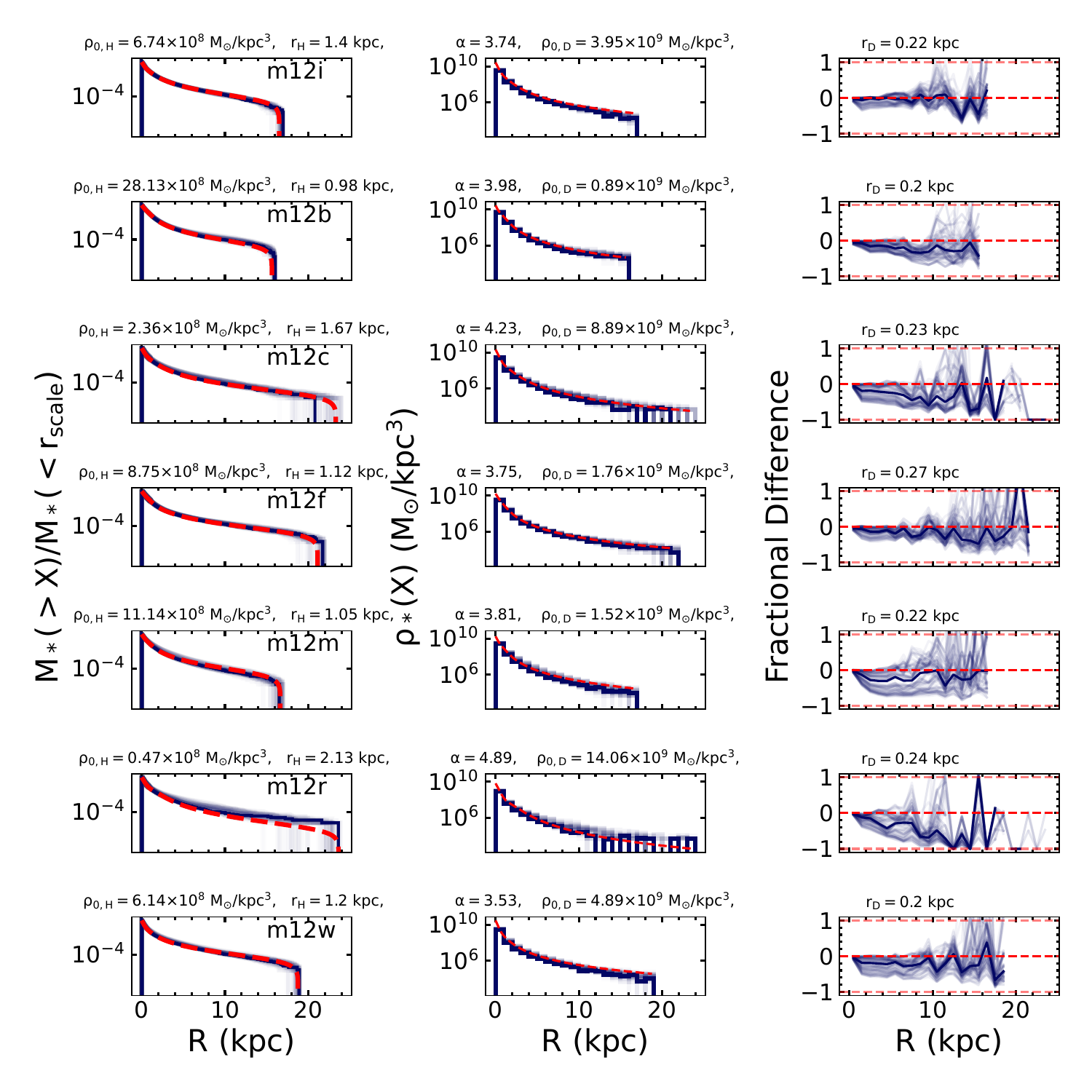}
    \caption{ The minor axis stellar density profile fits for the FIRE-2 isolated galaxies where each row demonstrates the fit for a different galaxy whose label is in the top right corner of the left panels. The CDF fits are shown in the left panels (red dashed line) with the galaxy's stellar particle CDF (dark blue solid line). In the left and middle panels, we also show 100 different stellar profiles for cylinders that are randomly selected to be up to 45$^{\circ}$ away from the minor (z-)axis (dark blue transparent lines). The middle panels shows the unnormalized stellar particle distributions. The final fit distribution is shown as a red dashed line in the middle panels, with the final parameters printed above the panels for each galaxy. In the right panel, we show the fractional difference between particle distribution along the z-axis compared to the 100 random directions (transparent blue lines). The median of these lines is shown as the dark blue solid line with no transparency. The red dashed line indicate the y-axis values of 0, -1 and 1. }
    \label{fig:si}
\end{figure}

\begin{figure}
    \centering
    \includegraphics[width=\linewidth]{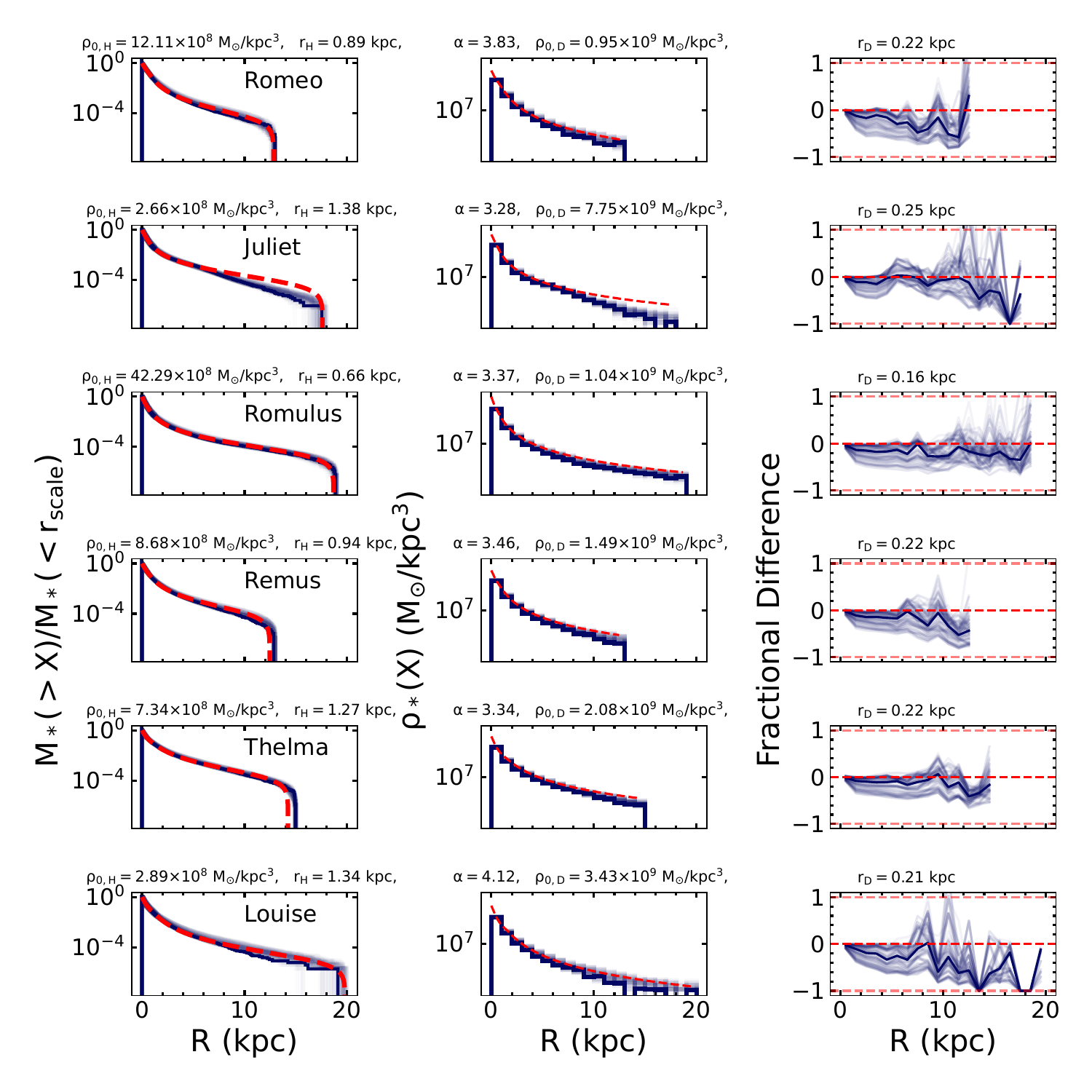}
    \caption{The same as Figure \ref{fig:si} but for the FIRE-2 galaxy pairs.}
    \label{fig:sp}
\end{figure}

\begin{figure}
    \centering
    \includegraphics[width=\linewidth]{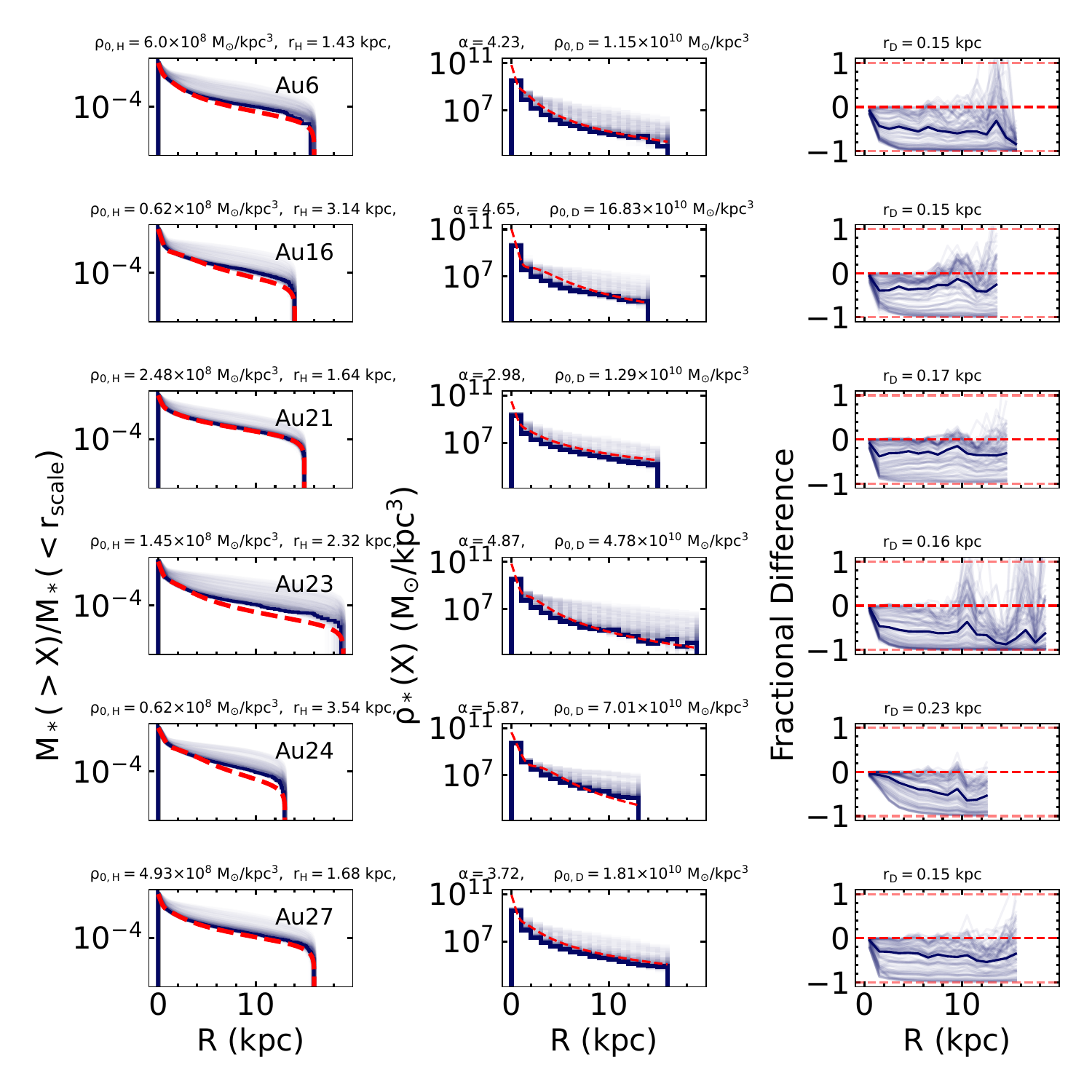}
    \caption{The same as Figure \ref{fig:si} but for the Auriga galaxies.}
    \label{fig:sa}
\end{figure}

\begin{figure}
    \centering
    \includegraphics[width=\linewidth]{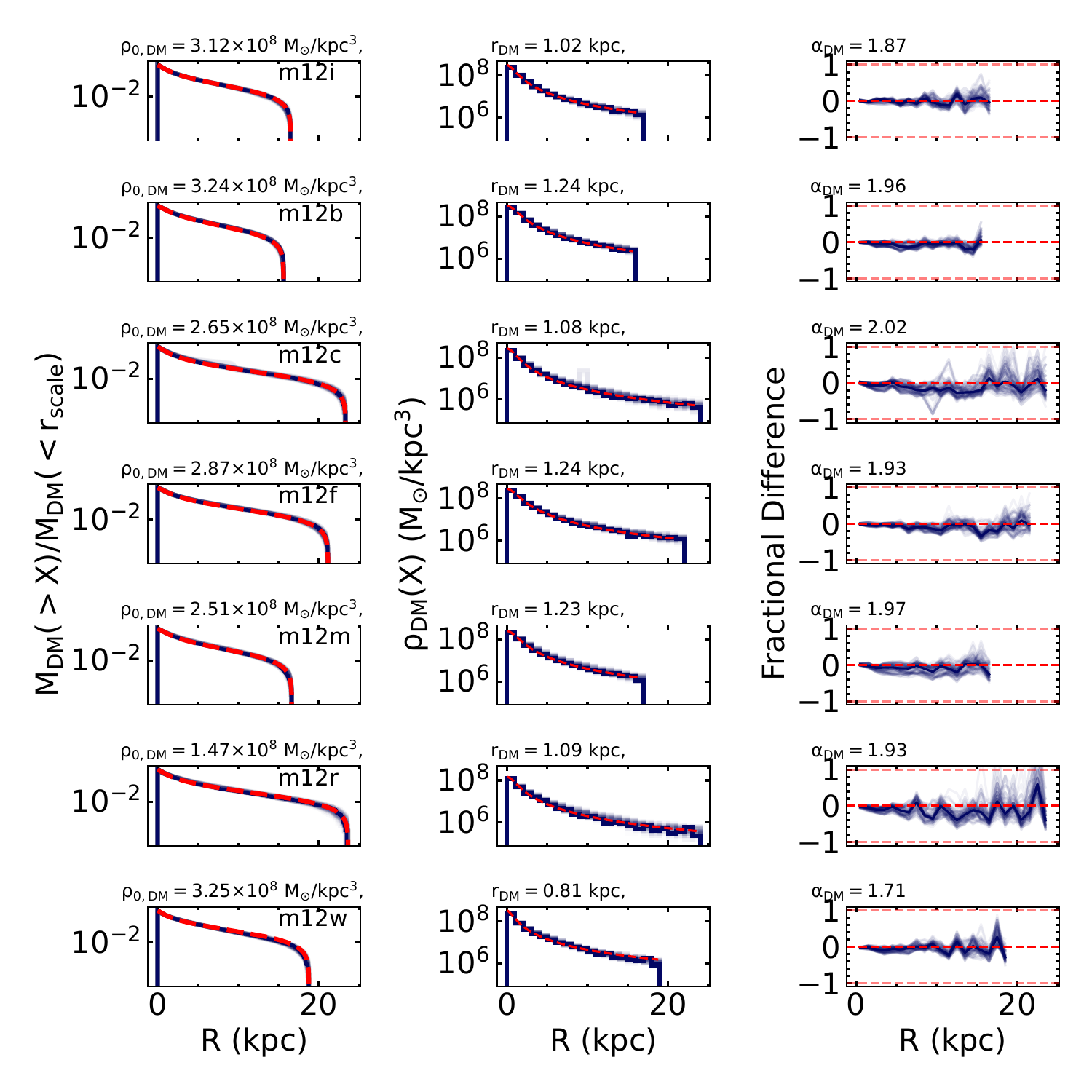}
    \caption{The same as Figure \ref{fig:si} but for the dark matter profiles.}
    \label{fig:di}
\end{figure}

\begin{figure}
    \centering
    \includegraphics[width=\linewidth]{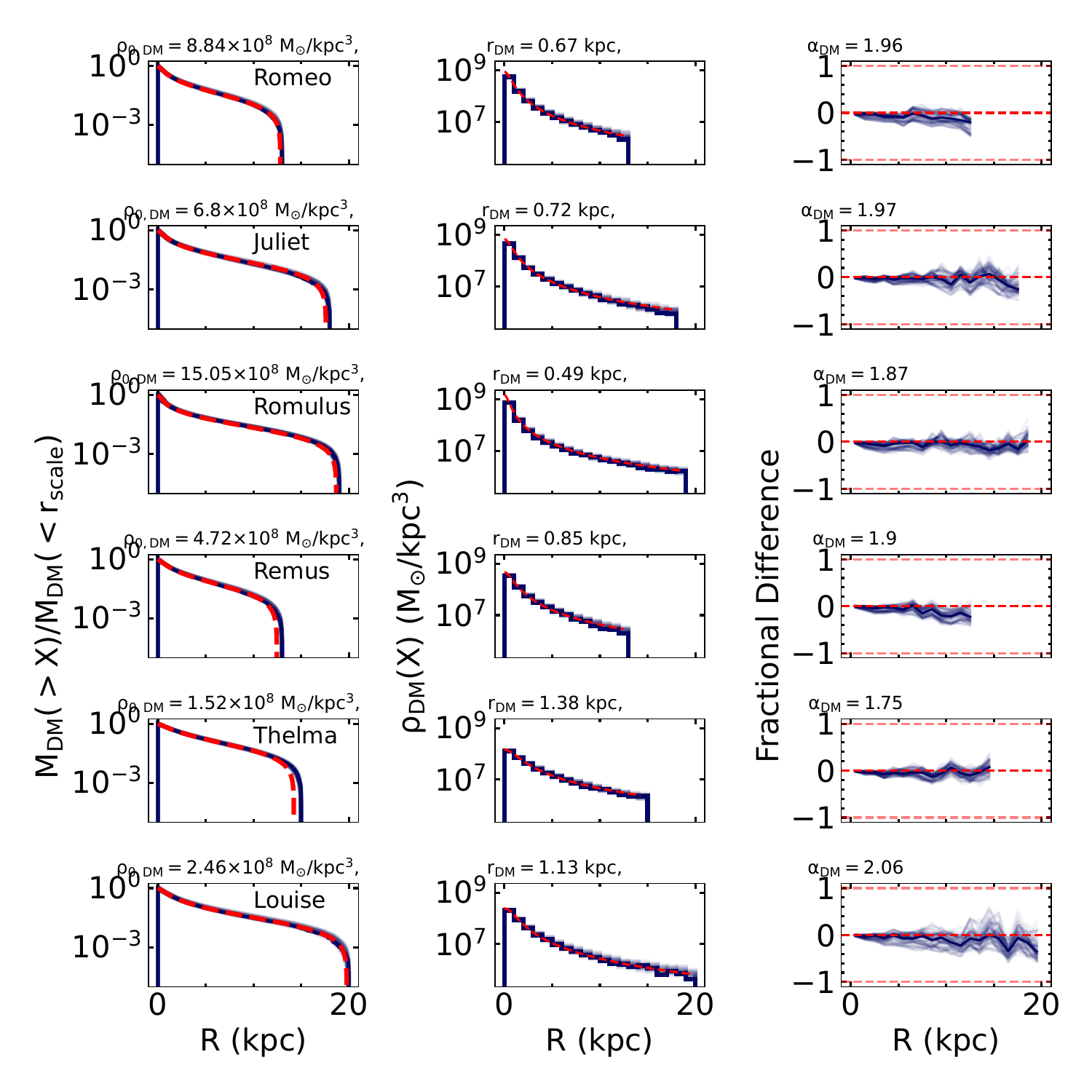}
    \caption{The same as Figure \ref{fig:di} but for the FIRE-2 galaxy pairs.}
    \label{fig:dp}
\end{figure}

\begin{figure}
    \centering
    \includegraphics[width=\linewidth]{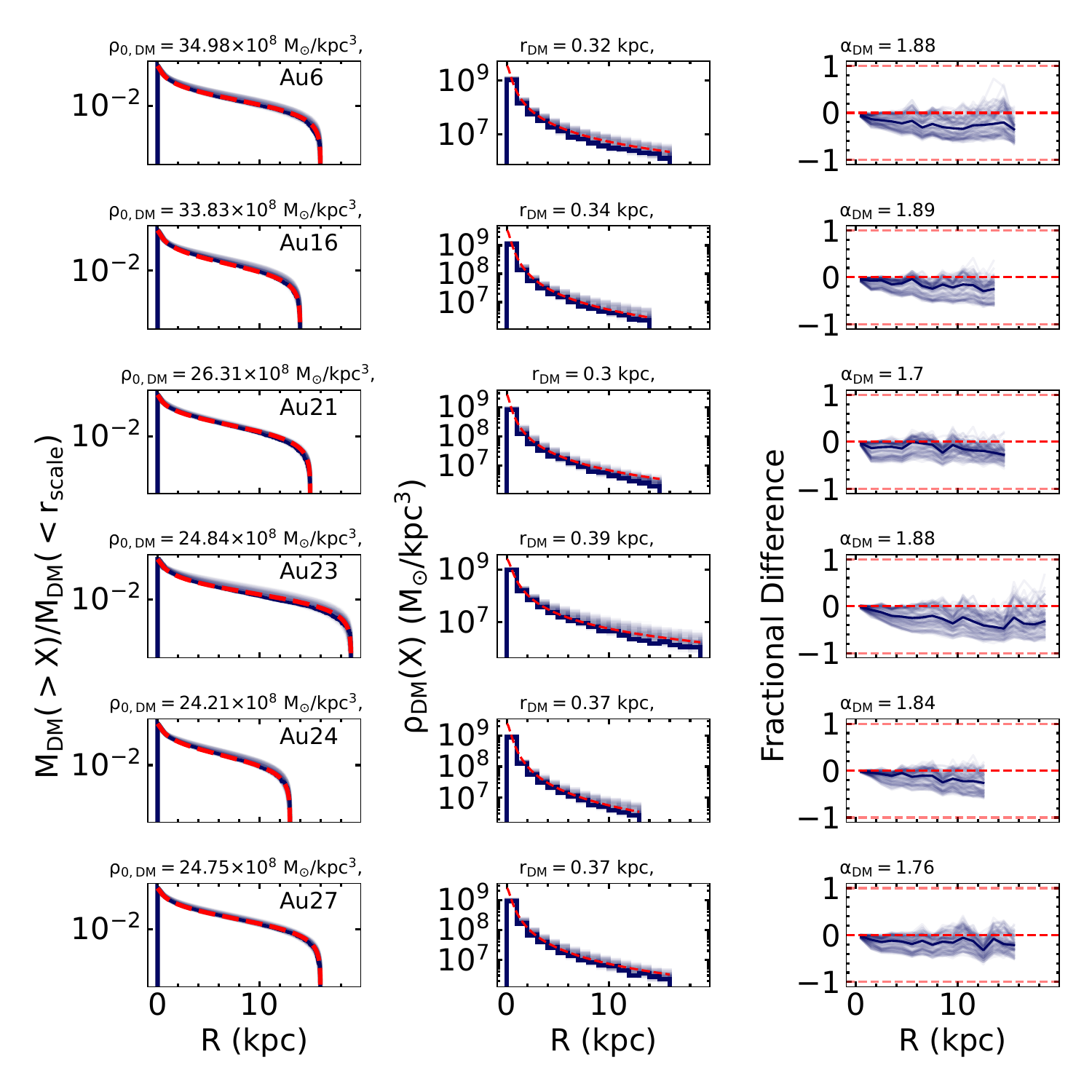}
    \caption{The same as Figure \ref{fig:di} but for the Auriga galaxies.}
    \label{fig:da}
\end{figure}

To parameterize the minor axis stellar and dark matter density profiles, we fit simple power law functions (see Equations 2 and 3) to the distribution of particles inside a cylinder of radius $1/\sqrt{\pi}$ along the minor axis which is defined to be perpendicular to the galaxy's disk. For the stellar component we also need to account for the exponential disk. To avoid the effect of binning, we first perform the fit to the cumulative distribution function which is normalized by the total stellar mass within the cylinder out to the halo's NFW scale radius (left panels).  To derive the fit parameters and corresponding uncertainties we use $scipy.optimize.curve\_fit$ where the covariance matrix is calculated using the linear approximation method.

In Figures \ref{fig:si}, \ref{fig:sp}, and \ref{fig:sa}, we show the fits to the stellar density profiles, while Figures \ref{fig:di}, \ref{fig:dp}, and \ref{fig:da} show the fits to the dark matter profiles. Each row corresponds to one simulated galaxy with the resulting parameterization printed above the corresponding panels. The galaxy's designation is printed in the top right of the leftmost panel. The middle panels show the fit (red dashed line) to the unnormalized density distribution (dark blue solid line). We note that fits to the Auriga stellar density profiles are generally worse then the fits to the FIRE-2 galaxies. For example, the $\rm{log_{10}}$ RMSE is 7.17 for the isolated FIRE-2 galaxies, 7.15 for the pairs and 7.38 for the Auriga galaxies. This is because the functional form (Equation 2) was chosen to optimize the fit to the FIRE-2 galaxies. The Auriga galaxies may be fit better using a different functional form, but in order to facilitate a comparison we keep the functional form the same for both simulations. 

In the left and middle panels, we also plot in transparent dark blue the stellar particle distributions  for cylinders of the same size, but at 100 random angles that are $<$ 45$^{\circ}$ from the z-axis. With these distributions, we investigate the level of substructure and asymmetry in the simulated minor axis profiles.  The right panels show the difference between the density distributions along the z-axis and the 100 random angles, normalized by the z-axis density at that point (dark blue transparent lines). The median of these 100 lines is shown as the solid dark blue line. We also show three red horizontal dashed lines at y-values of -1, 0, and 1. Generally, we find the distributions do not vary significantly for the FIRE-2 galaxies, while the Auriga galaxies show systematically larger densities at angles closer to the disk, indicating more oblate halos, consistent with results from \citet{Monachesi2019}.


\bibliography{bibliography}{}
\bibliographystyle{aasjournal}



\end{document}